\documentclass[reprint,amssymb,noeprint,twocolumn,longbibliography]{revtex4-2}

\usepackage{graphicx}
\usepackage{amsfonts}
\usepackage{amsmath}

\usepackage[usenames,dvipsnames]{color}
\usepackage{bm}
\usepackage{amssymb}
\usepackage{braket}

\usepackage{natbib}

\usepackage{multirow}
\usepackage{epstopdf}
\usepackage[normalem]{ulem}

\usepackage{hyperref}
\hypersetup{
colorlinks=true,
linkcolor=blue,
citecolor=blue,
filecolor=green,
urlcolor=blue,
}
\usepackage[nameinlink,capitalise]{cleveref}

\begin{document}


\title{Magnetically tunable electric dipolar interactions of ultracold polar molecules in the quantum ergodic regime}


\author{Rebekah Hermsmeier$^{1}$, Ana Maria Rey$^{2}$, and Timur V. Tscherbul$^{1}$}

\affiliation{$^{1}$Department of Physics, University of Nevada, Reno, Nevada, 89557, USA}
\affiliation{$^{2}$JILA, National Institute of Standards and Technology, and Department of Physics, University of Colorado,  Boulder, Colorado, 80309, USA}

\date{\today}

\begin{abstract}
By leveraging the hyperfine interaction between the rotational and nuclear spin degrees of freedom, we demonstrate extensive magnetic control over the electric dipole moments, electric dipolar interactions, and ac Stark shifts of  ground-state alkali-dimer molecules such as KRb$(X^1\Sigma)$. The control is enabled by narrow avoided crossings and the highly ergodic character of molecular eigenstates at low magnetic fields, offering a general and robust way of continuously tuning  the intermolecular electric dipolar interaction for applications in quantum simulation and sensing.

 \end{abstract}
\maketitle
\newpage

 The nearly limitless availability of quantum levels with long lifetimes, favorable coherence properties, and strong, tunable electric dipolar (ED) interactions make ultracold polar molecules a highly attractive platform for quantum science  \cite{DeMille:02,Yelin:06,Yan:13,Bohn:17,Albert:20,Park:17,Li:23,Tobias:22,Christakis:23,Zhang:20,He:20,Cairncross:21,zhang2022optical,Burchesky:21,holland2022demand,bao2022dipolar}, ultracold chemistry \cite{Krems:08,Balakrishnan:16}, and precision searches of new physics beyond the Standard model \cite{Bohn:17, DeMille:17}.
 Attaining  robust quantum control over molecular electric dipole moments (EDMs) and their ED interactions is key to achieving high-fidelity quantum gates  \cite{Yelin:06,Ni:18} and dynamical generation of entangled states   \cite{bao2022dipolar,holland2022demand}, which can be used for a wide range of applications ranging from quantum metrology  \cite{Pezze:18,Bilitewski:21,Tscherbul:23,Bilitewski:23} 
to quantum simulation  \cite{micheli2006toolbox,Micheli:07,Carr:09,Gorshkov:11b,Hazzard:13,Yan:13}.

Thus far, quantum control of EDMs and ED interactions of  polar molecules has only been explored for internal molecular states in the high magnetic ($B$) field limit, where the nuclear spins are nearly completely polarized and decoupled from molecular rotation, leading to magnetic field-insensitive molecular eigenstates of the form $\ket{NM_N}\ket{I_1M_{I_1}}\ket{I_2 M_{I_2}}$, where $\hat{\mathbf{N}}$
 is the rotational angular momentum of the molecule with eigenvalue $\sqrt{N(N+1)}$, $\hat{\mathbf{I}}_m$ are the spin operators for the $m$-th nucleus ($m=1,2$), and $M_N$, and $M_{I_m}$ are the projections of $\hat{\mathbf{N}}$ and $\hat{\mathbf{I}}_m$ on the $B$-field axis.  As a result, in this limit, $B$ fields cannot be used  to tune the EDMs of closed-shell polar molecules (such as KRb, NaK, and NaCs), precluding magnetic control over their ED interactions.
While these interactions can still be tuned using external dc and ac electric ($E$) fields, this type of control imposes a number of significant  limitations.
For example, it is impossible to turn the ED interaction  off on demand using a dc $E$ field alone when working with   superpositions of  rotational states with non-zero ED coupling, such as $N=0$ and $N=1$ states. Thus far, this challenge has been addressed by transferring the molecules to (or from) the states coupled (decoupled) by the ED interaction \cite{Ni:18,Yan:13}, requiring additional  microwave or Raman pulses.


Here, we propose a general mechanism for smoothly tuning the ED interactions between polar molecules via  an external dc magnetic field.  By leveraging the hyperfine interactions between the nuclear spin and rotational degrees of freedom  in polar alkali-dimers, we demonstrate magnetic control over their EDMs  over a wide dynamic range enabled by the ergodic behavior of molecular eigenstates at  low $B$ fields. Unlike previous work \cite{Gorshkov:11b,Tscherbul:23}, our proposal does not require strong magnetic fields and/or microwave dressing, and can be realized with molecules remaining in a single quantum state, thus obviating the need for coherent state transfer to switch the ED interaction on and off \cite{Yelin:06,Ni:18}.
Our results thus demonstrate the possibility of on-demand generation and tuning of long-range ED interactions in ultracold molecular gases using dc magnetic fields alone. 


We start by considering the energy level spectrum of an alkali dimer molecule (e.g., KRb) in its ground electronic and vibrational states 
described by the Hamiltonian \cite{Aldegunde:08}
\begin{equation}\label{eq:Hmol}
\begin{aligned}
\hat{H}_\text{mol}=\hat{H}_\text{rot}+\hat{H}_{hf}+ \hat{H}_Z+ \hat{H}_S,
\end{aligned}
\end{equation} 
where $\hat{H}_\text{rot}=B_e  \hat{\textbf{N}}^2$ is the rotational Hamiltonian, $\hat{\textbf{N}}$ is the 
rotational angular momentum operator and $B_e$ is the rotational constant.  The Stark Hamiltonian $\hat{H}_S=-\hat{\textbf{d}} \cdot \hat{\textbf{E}}$ describes the interaction of the EDM $\mathbf{d}$ with a dc electric field $\mathbf{E}$ directed along a space-fixed (SF) $z$ axis.  The  interaction of the molecule with an external magnetic field $B=|\mathbf{B}|$, parallel to the $E$ field, is given by 
$\hat{H}_Z=- g_N \mu_N \hat{{N}}_z B- \sum_{m=1,2} g_{I_m} \mu_N \hat{{I}}_{m_z} B (1-\sigma_m)$, 
where $g_N$ is the rotational $g$-factor, $g_{I_m}$ is the $g$-factor for the $m$-th nucleus, 
$\mu_N$ is the nuclear Bohr magneton, and $\sigma_m$ are the shielding factors. The hyperfine Hamiltonian $\hat{H}_{hf}$  
\begin{equation}\label{eq:Hhf}
c_3 \hat{\textbf{I}}_1 \cdot \hat{\textbf{I}}_2+\sum_{m,p} (-1)^p C_p^2(\theta,\phi) \frac{\sqrt{6} (eqQ)_m}{4 I_m(2I_m-1)}
 T^2_{-p}(\hat{\textbf{I}}_m,\hat{\textbf{I}}_m),
\end{equation}
is dominated by the scalar spin-spin interaction $c_3 \hat{\textbf{I}}_1 \cdot \hat{\textbf{I}}_2$ and the nuclear electric quadrupole (NEQ) interaction  \cite{Brown:03,Aldegunde:08}. The latter arises from the non-spherical shape of the $I\ge1$ nuclei in KRb leading to nonzero NEQ moments, which interact with the (non-spherical) electron charge distribution inside the molecule. The interaction energy depends on the orientation of the NEQ ellipsoid (whose axis of symmetry coincides with the direction of $\mathbf{I}_m$) with respect to the molecular axis $\mathbf{r}$ \cite{Fano:48}. Transforming this electrostatic interaction to the SF frame, 
one obtains the second term in Eq.~\eqref{eq:Hhf}, where $C^k_p(\theta,\phi)$ is a renormalized spherical harmonic, $\theta$ and $\phi$ specify the orientation of $\mathbf{r}$ in the SF frame \cite{Zare:88}, $(eQq)_m$ are the NEQ interaction constants, and  $T^2(\hat{\textbf{I}}_m,\hat{\textbf{I}}_m)$ is the second-rank tensor  product of $\hat{\textbf{I}}_m$ with itself \cite{Brown:03}.  The NEQ interaction couples the bare states $|N M_N, M_{I_1} M_{I_2} \rangle$ with $|N-N'|\leq 2$ and $|M_N-M_N'|\leq 2$ \cite{Gorshkov:11b} but vanishes in the $N=0$ manifold. As a result, in the previously underexplored low $B$ field regime of interest here, the eigenstates in the $N=1$ manifold contain contributions from many bare states with different $M_N$, $M_{I_1}$, and $M_{I_2}$.

\begin{figure}
 \centering
 \includegraphics[width=1.02\columnwidth,trim = 0 0 0 10]{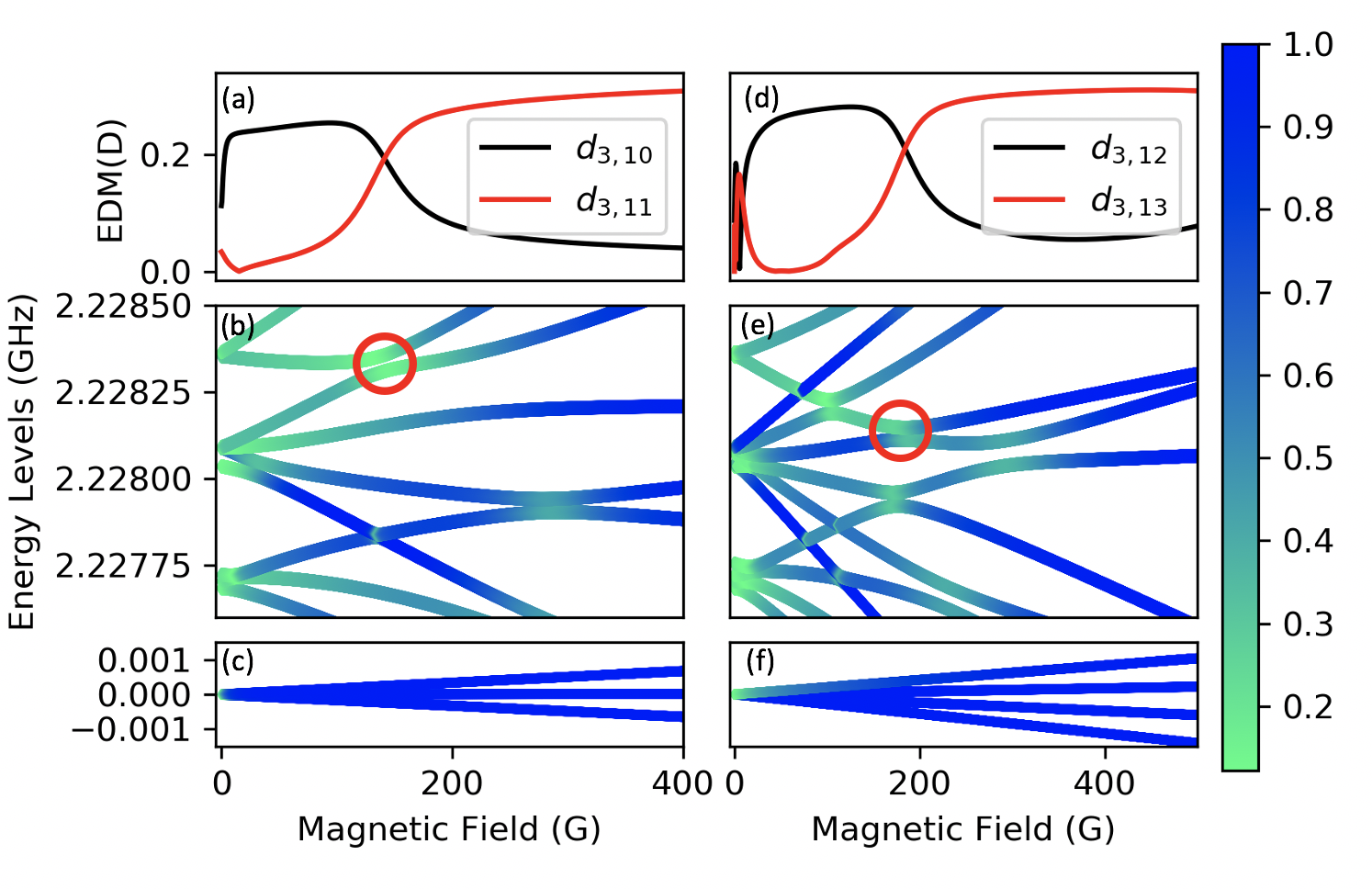}
 \caption{Representative transition EDMs $d_{3,10}$ and $d_{3,11}$ of $^{40}$K$^{87}$Rb plotted as a function of magnetic field at zero electric field for $M_F= 7/2$ (a) and  $M_F=-3/2$ (d). The corresponding $N=0$ and $N=1$ hyperfine-Zeeman energy levels are shown in panel (c) and (b) respectively.  The color of the  energy levels corresponds to the IPR ergodicity measure ${P}_\alpha(i|i)$ of the corresponding energy eigenstates. Encircled are the avoided crossings between the $N=1$ states $\ket{10}$ and $\ket{11}$ [panel (b)] and $\ket{12}$ and $\ket{13}$  [panel (e)] responsible for the switching behavior of the EDMs in panels (a) and (d).  }
  \label{fig:EnergyLevelEDM}
\end{figure}


 Figure \ref{fig:EnergyLevelEDM}  shows the energy levels and transition EDMs of KRb  as a function of magnetic field at $E=0$ obtained by exact diagonalization of the Hamiltonian \eqref{eq:Hmol} 
 \cite{Gorshkov:11b,Tscherbul:23}. We label the levels in the order of increasing energy (with $\ket{1}$ being the ground state). The nuclear spins of K and Rb are $I_1=4$ and $I_2=3/2$, giving rise to $(2I_1+1)(2I_2+1)(2N+1)$ or 36 (108) Zeeman sublevels in the $N=0$ ($N=1$) rotational manifolds \cite{Aldegunde:08}. The total angular momentum projection along the applied electromagnetic fields, $M_F=M_N+M_{I_1}+M_{I_2}$,  where 
 $\hat{\mathbf{F}}=\hat{\mathbf{N}} + \hat{\mathbf{I}}$ 
 ($\hat{\mathbf{I}}=\hat{\mathbf{I}}_1 + \hat{\mathbf{I}}_2$) is conserved in parallel $E$ and $B$ fields, so only a smaller subset of levels (3 for $N=0$  and 9 for $N=1$) occurs in the $M_F=7/2$ symmetry sector \cite{Aldegunde:08,Aldegunde:09}. 
 In the $N=0$ manifold at $B\leq 20$~G, the scalar spin-spin interaction splits the energy levels into four manifolds $F=\{11/2,9/2,7/2,5/2\}$ according to the value of the total angular momentum $F$,  which ranges from $|I - N|$ to $|I+N|$. In the $N \ge 1$ manifolds the dominant effect at low $B$ fields is the NEQ interaction, which  splits the energy levels by their $\hat{\mathbf{F}}_2=\hat{\mathbf{N}} +\hat{\mathbf{I}}_2$ angular momenta values, $F_2=5/2,3/2,1/2$, where $F_2=|I_2-N| \leq F_2\leq I_2+N$. In the high $B$ field limit, the Zeeman interaction takes over and groups the eigenstates by the values  of good quantum numbers $M_N, M_{I_1},$ and $M_{I_2}$.
 
 Remarkably, as shown in Fig.~\ref{fig:EnergyLevelEDM}(a), the absolute magnitude of the matrix element {$d_{3,11}=|\langle 3| \hat{d}_0|11\rangle|$ } increases sharply from zero to a nearly constant value, while $d_{3,10}$  decreases to zero at high $B$ fields.
To explain these variations, we notice that there is an avoided crossing between the energy eigenstates $|10\rangle$ and $|11\rangle$ at the same  field ($140$~G), where  $d_{3,10}$ and $d_{3,11}$ swap values.  
The crossing causes the eigenstates $|10\rangle$ and $|11\rangle$ to switch their dominant bare states, only one of which has nonzero transition EDM with the ground state $|3\rangle = |00, 4-\frac{1}{2}\rangle$, which explains the switching.  
Indeed, to the  left of the avoided crossing encircled in Fig.~\ref{fig:EnergyLevelEDM}(b), $|10\rangle \simeq |10,4-\frac{1}{2}\rangle$ and $|11\rangle\simeq |11,3-\frac{1}{2}\rangle$ and thus  $d_{3,10}\to \text{const}$  and $d_{3,11}\to 0$.
 To the right of the crossing, the bare states are swapped, and hence $d_{3,10}\to \text{const}$  and $d_{3,11}\to 0$ at large $B$.
 This tunability is ubiquitous at low $B$ fields as illustrated in Fig.~\ref{fig:EnergyLevelEDM}(d) (see also the Supplemental Material). 

  


At a more quantitative level, the $B$ field dependence of the EDM matrix elements can be described by expanding  molecular eigenstates $\ket{i}$ in bare states  $\ket{\alpha}=|NM_N, M_{I_1}M_{I_2}\rangle$ as 
\begin{equation}\label{eigenstate_expansion}
\ket{i}=\sum_{\alpha} c_{\alpha,i}(B)\ket{\alpha}.
\end{equation}
Because the  EDM operator $\hat{d}_0=\hat{d}_z$ is diagonal in $M_{I_1}$ and $M_{I_2}$, its matrix elements in the eigenstate basis are ($\delta_{MM'}=\delta_{M_{I_1}M_{I_1'}}\delta_{M_{I_2}M_{I_2'}}$)
\begin{equation}\label{d0_general}
\langle i|\hat{d}_0|j\rangle=
\sum_{\alpha,\alpha'}
c_{\alpha, i}(B)  
c_{ \alpha' , j}(B) \delta_{MM'}
\langle N M_N |  \hat{d}_0| N’ M’_N \rangle.
\end{equation} 
Significantly, because the matrix elements on the right-hand side are independent of $B$, the magnetic tunability of the EDM  arises entirely from the expansion coefficients  $c_{\alpha, i}(B)$, which depend on the extent of ergodicity of molecular eigenstates (see below).
 Using the conservation of $M_F$ to narrow down the range of states, which contribute to Eq.~\eqref{d0_general}, and noting that $\ket{3}\to|00,4-\frac{1}{2}\rangle$ above 30~G, we obtain $d_{3,11}=|b_1(B) \langle 00 |\hat{d}_0|10\rangle|$, where $b_1=c_{10 4-\frac{1}{2},11}(B)$.
The $B$ field dependence of the transition EDM shown in Fig.~1(a) is then completely determined by that of  $b_1$,
as can be seen by comparing  Figs.~2(a) and 2(b).
   Figure~2(a) shows that  the prediction of the single-term model is in excellent agreement with the numerically exact  transition EDM.  

\begin{figure}[]
 \centering
 \includegraphics[width=0.8\columnwidth, trim = 0 0 0 -5]{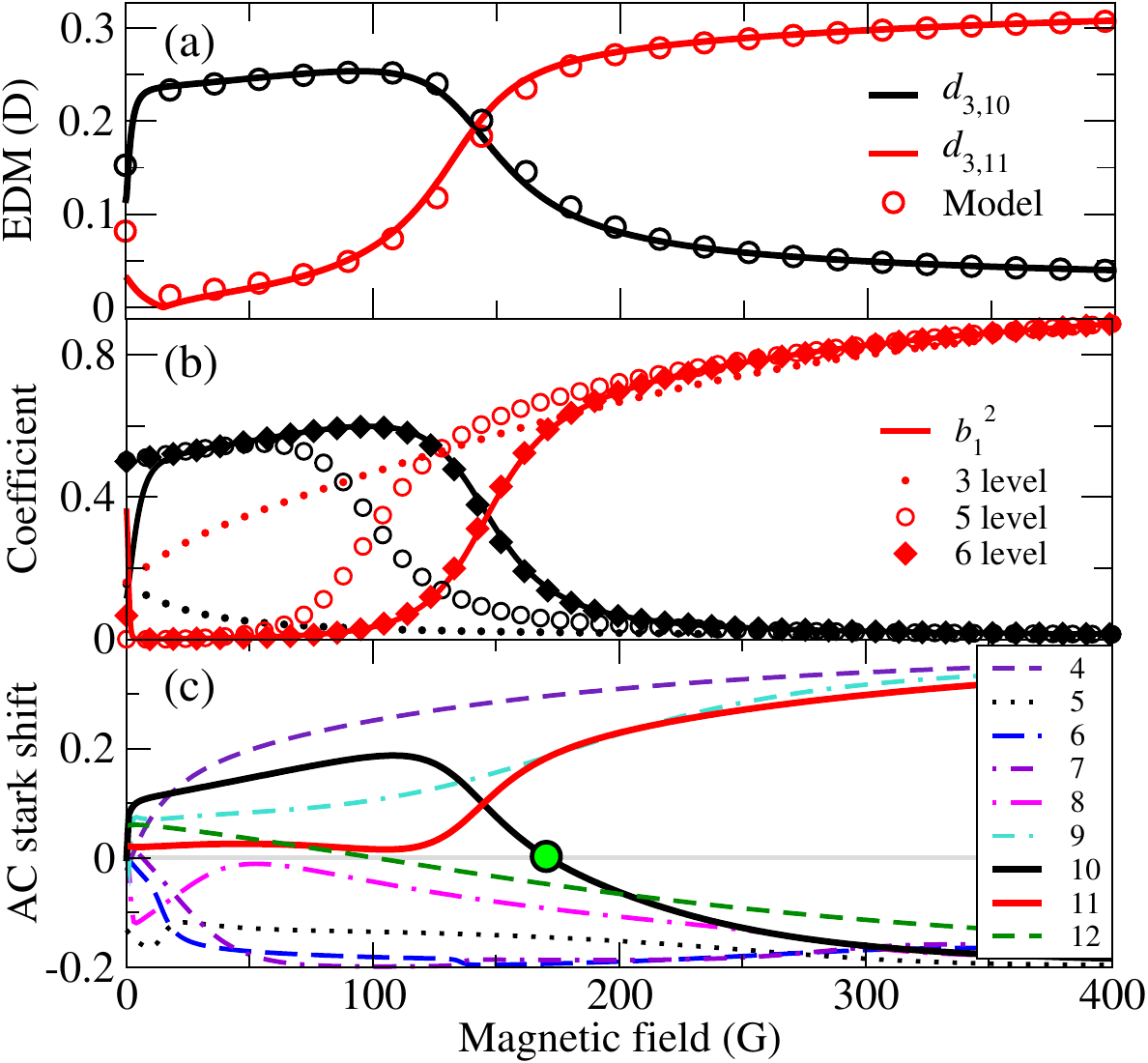}
 \caption{(a) Transition EDMs  $d_{3,10}$ and $d_{3,11}$ plotted as a function of $B$ field at $E=0$ and $M_F=7/2$. Full lines: exact result, symbols: single-state model. (b) The dependence $b_1(B)^2$. Full lines: exact result, symbols: model results including 3, 5, and 6 bare states.  (c) ac Stark shifts for the states shown in Fig.~1(b) as a function of $B$ field ($E=0$).  The green circle marks the magic $B$ field for state $\ket{10}$.
 }
  \label{fig:EDMcoefficients}
\end{figure}

As the $B$ field decreases below $200$ G, the amplitude $b_1(B)$ declines monotonically from 1 [see Fig. 2(b)] as the NEQ interaction begins to admix other bare states into eigenstate $\ket{11}$ such as {$|11,3-\frac{1}{2}\rangle$, $|1-1,3\frac{3}{2}\rangle$, and $|10,3\frac{1}{2}\rangle$. These bare states ``dilute'' the eigenstate, 
causing the transition EDM $d_{3,11}$ to decline to zero. As shown in Fig.~2(b) it is necessary to retain as many as 6 bare states in Eq.~\eqref{eigenstate_expansion} to reproduce the field dependence of $b_1(B)$.
The single-term model also explains the behavior of the transition EDM $d_{3,10}\simeq  |b_2(B) \langle 00| \hat{d}_0| 10 \rangle|$, where $b_2(B)=c_{104-\frac{1}{2},10}(B)\to 0$ at large $B$, as shown in Fig.~2(b) due to the eigenstate $|10\rangle$ approaching the bare state $|10,3\frac{1}{2}\rangle$.

To further elucidate the $B$ field dependence of the EDMs, we invoke the concept of quantum ergodicity \cite{Nordholm:74,Pittman:15}, which plays a central role in theoretical studies of intermolecular vibrational energy redistribution \cite{Uzer:91,Bigwood:98,Nesbitt:96,Leitner:15}. Very recently, ergodicity-breaking  transitions have been observed in C$_{60}$ molecules  as a function of  rotational angular momentum \cite{Liu:23}.  An eigenstate $\ket{i}$ is said to be ergodic with respect to a bare (or zeroth-order) basis set  $\{\ket{\alpha}\}$ if it is delocalized in the Hilbert space spanned by  $\{\ket{\alpha}\}$. The degree of ergodicity of a given molecular eigenstate  can be quantified \cite{Pittman:15} by the  inverse population ratio  (IPR) 
${P}_{\alpha}(i|i)= \sum_\alpha |c_{\alpha,i}|^4$, 
where $c_{\alpha,i}$ are defined by Eq.~\eqref{eigenstate_expansion}.
A small value of ${P}_{\alpha}(i|i)$ indicates that $\ket{i}$ is highly mixed with respect to the bare state basis
 $\ket{\alpha}=|NM_N, M_{I_1}M_{I_2}\rangle$, which gives a physically meaningful representation of the EDM operator.

Figures~\ref{fig:EnergyLevelEDM}(b)-(f) show the IPR for the lowest eigenstates of KRb as a function of magnetic field. In the high $B$-field limit the eigenstates consist primarily of a single bare state, and thus {${P}_{\alpha}(i|i)\simeq 1$}. As we reach lower $B$ fields, the NEQ coupling causes different bare state contributions to the individual eigenstates to mix, increasing  their ergodicity and lowering ${P}_{\alpha}(i|i)$. In addition, we observe larger ergodicity near  avoided crossings of the $N=1$ levels, which reflects additional NEQ mixing of the nearly degenerate bare states.


 \begin{figure}[]
 \centering
 \includegraphics[width=1.05\columnwidth]{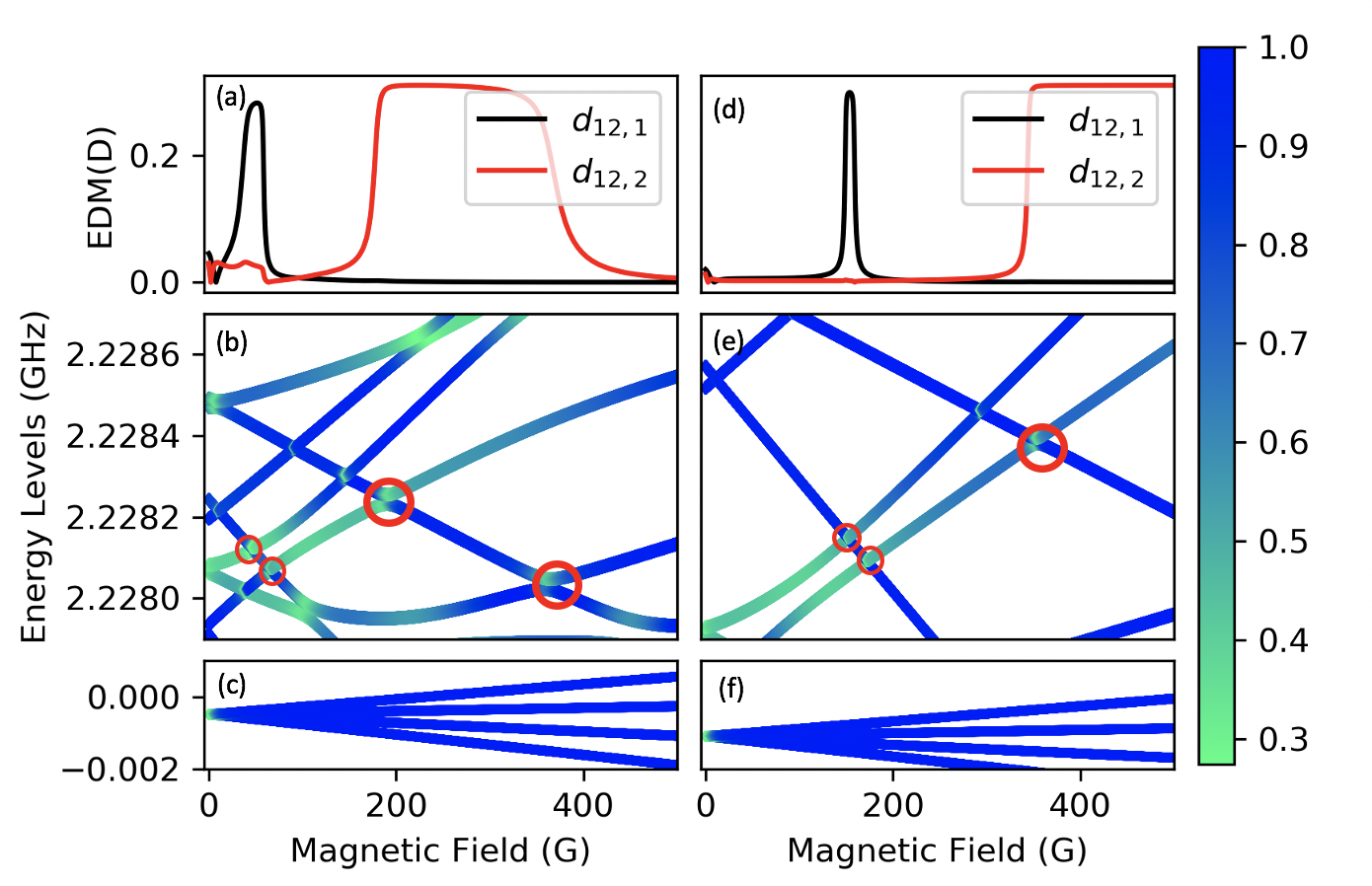}
 \caption{Representative transition EDMs $\langle 12|\hat{d}_0|1\rangle$ and $\langle 12|\hat{d}_0|2\rangle$ of KRb plotted as a function of magnetic field  for $M_F= -3/2$ at $E=0.2$ kV/cm (a) and  $E=0.3$ kV/cm (d). The corresponding $N=0$ and $N=1$ Zeeman energy levels are shown in panel (c) and (b) respectively.  The color of the  energy levels corresponds to the ergodicity of molecular eigenstates. Encircled are the avoided crossings between the $N=1$ levels responsible for the tunable behavior of the EDMs in panels (a) and (d). }
  \label{fig:EDMElectricEnergyLevels}
\end{figure}

To test whether the exquisite tunability of the EDMs in the quantum  ergodic regime extends to other observables, we consider tensor ac Stark shifts $\Delta E^\text{ac}_i$ of the hyperfine sublevels $\ket{i}$ induced by the optical fields used in trapping experiments \cite{Kotochigova:10b,Neyenhuis:12,Blackmore:18,Sesselberg:18,Gregory:21,Lin:22}.  Minimizing these shifts is crucial for achieving long coherence times of ultracold molecules trapped in optical lattices and tweezers \cite{Yan:13,Bohn:17,Li:23,Tobias:22,Christakis:23,Zhang:20,He:20,Cairncross:21,zhang2022optical,Burchesky:21,holland2022demand,bao2022dipolar}. Assuming that the optical trapping field is off-resonant and sufficiently weak, one can show that $\Delta E^\text{ac}_i\simeq \langle i|P_2(\cos\theta)|i\rangle$, where $P_2(\cos\theta)$ is a second-order Legendre polynomial, which represents the anisotropic ($N$-dependent) part of molecular polarizability \cite{Kotochigova:10b,Gorshkov:11b}.
Figure~\ref{fig:EDMcoefficients}(c) shows that the tensor ac Stark shifts of the $N=1$ hyperfine sublevels of KRb can be efficiently controlled by applying a moderate $B$ field.  The shifts follow the same trends as those displayed by transition EDMs, showing rapid variations near avoided crossings and at low $B$ fields. In particular,  the ac Stark shifts of the states $\ket{10}$ and $\ket{11}$ show the same behavior as transition EDMs in Fig.~\ref{fig:EDMcoefficients}(a), which can be explained as described above.  Importantly, we observe that  tensor ac Stark shifts vanish at certain ``magic'' $B$-fields [see Fig.~\ref{fig:EDMcoefficients}(c)]. This shows that, similarly to $E$ fields \cite{Kotochigova:10b,Neyenhuis:12}, $B$ fields can be used  to prolong the coherence lifetimes of trapped alkali-dimer molecules.


Figures 3(a) and (d) show the magnetic field dependence of transition EDMs $d_{1,11}$ and $d_{2,12}$ in the presence of a small dc $E$ field.
In contrast to the zero $E$-field case, the EDMs display narrow peaks due to the additional avoided crossings seen in Figs.~3 (b) and (e). We also observe a decrease in the ergodicity of the eigenstates compared to the field-free case. This is caused by the Stark splitting between the $M_N=0$ and $M_N=\pm1$  levels, which  weakens the ergodicity-inducing NEQ coupling between these levels. As the $E$ field is further increased above a few kV/cm, we observe an increase in ergodicity due to the Stark coupling between the $N=0$ and $N=1$ states.
 In regions of low ergodicity, the eigenstates consist mainly of a single bare-state component. Thus, when an avoided crossing causes the eigenstates to switch their dominant bare-state components, the change in the eigenstate composition is significantly more dramatic than in the zero $E$-field case, causing rapid variations in the EDMs, as shown  in Figs.~3(a) and (d).


We now explore magnetic control of ED interactions between two polar alkali-dimer molecules trapped in an optical lattice or a tweezer as recently demonstrated experimentally \cite{Li:23,Tobias:22,Christakis:23,Zhang:20,He:20,Cairncross:21,zhang2022optical,Burchesky:21,holland2022demand,bao2022dipolar}. We encode an effective spin-$\frac{1}{2}$ system with eigenstates $|$$\uparrow \rangle$  and  $|$$\downarrow \rangle$ into molecular states chosen from the hyperfine-Zeeman levels in the $N=1$ and $N=0$ manifolds [see, e.g., Fig.~2(a)].  The ED interaction between the molecules $i$ and $j$ treated as effective spin-$\frac{1}{2}$  systems 
may be written as \cite{Wall:15c}
\begin{equation}\label{H_XXZ}
\hat{H}_{ij}=\frac{1-3 \cos^2\theta_{ij}}{|\textbf{R}_{ij}|^3}\bigg[\frac{J_\perp}{2} (\hat{S}^i_+  \hat{S}^j_- + \hat{S}^j_+ \hat{S}^i_-)+J_z \hat{S}^i_z \hat{S}^j_z\bigg],
\end{equation} 
where $\hat{S}^i_{\pm}$ and $\hat{S}^i_z$ are the effective spin-1/2 operators, $\theta_{ij}$ is the angle between $\mathbf{E}$ and the vector joining the molecules $\mathbf{R}_{ij}$, and $J_z=(d_{\uparrow} - d_{\downarrow})^2$ and $J_\perp= 2 d_{\uparrow\downarrow}^2$  are the Ising and spin-exchange coupling constants \cite{Wall:15c,Gorshkov:11b}. 
  The tunability of these constants is key to generating metrologically useful many-body entangled states  \cite{Perlin:20,Tscherbul:23} and exploring new regimes of far-from-equilibrium quantum magnetism \cite{Hazzard:13,Yan:13} using the XXZ Hamiltonian \eqref{H_XXZ}.

\begin{figure}[]
 \centering
 \includegraphics[width=\columnwidth, trim = 0 5 0 -40]{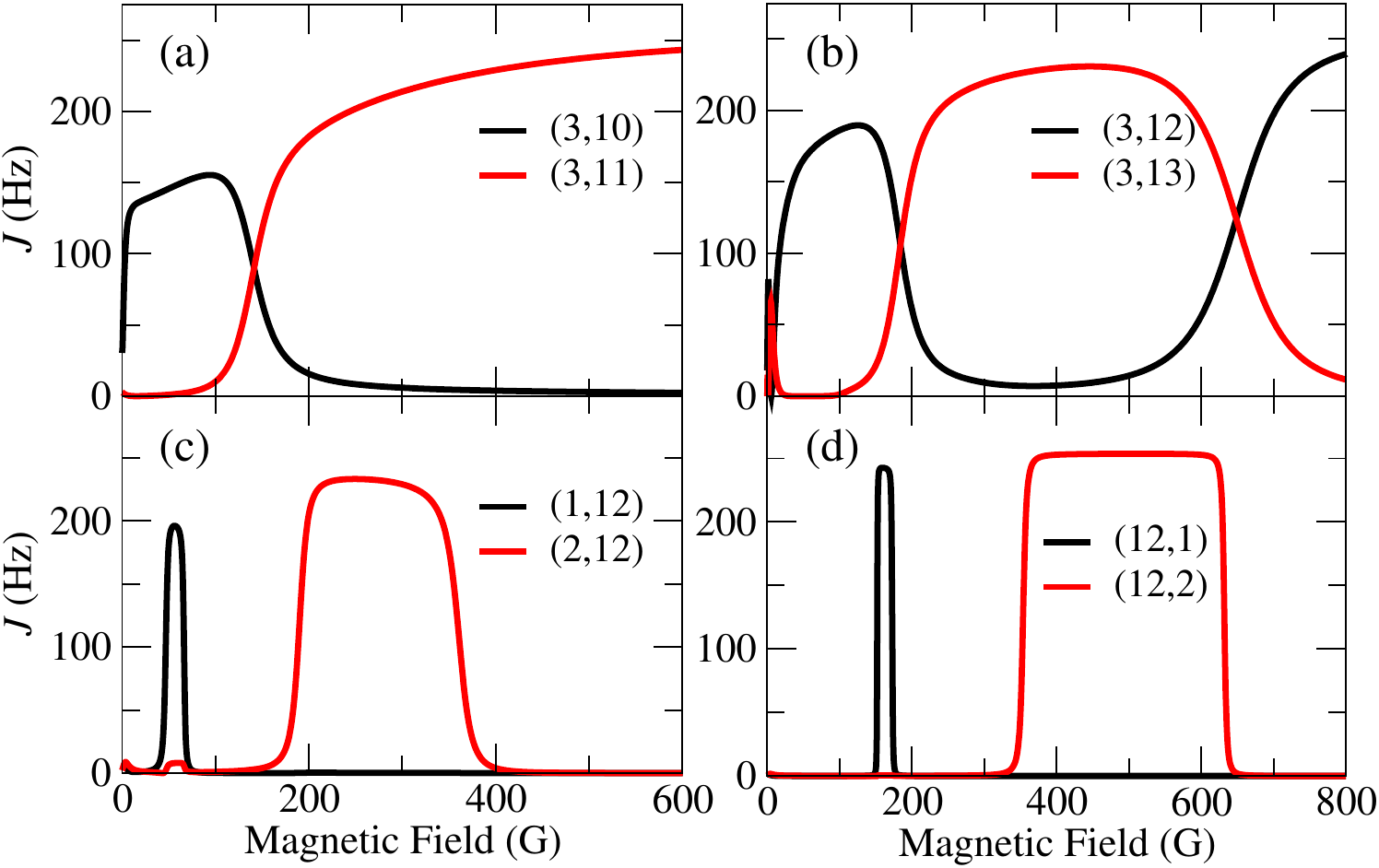}
 \caption{Spin exchange coupling constants $J_\perp$ plotted as a function of magnetic field at $E=0$ for $M_F= 7/2$ (a) and $M_F=-3/2$ (b).  Bottom panels show $J_\perp$  vs. $B$ for $M_F=-3/2$ at $E= 0.2$~kV/cm (c) and 0.3 kV/cm (d).}
  \label{fig:Jperp}
\end{figure}

Figure~4 shows the magnetic field dependence of the spin-exchange coupling $J_\perp$ for several representative encodings of the effective spin-1/2 into molecular hyperfine states, such as $\ket{\uparrow} =|3\rangle$,  $\ket{\downarrow } = |10\rangle$. Remarkably, we observe a strong variation of $J_\perp$ over a wide dynamic range ($0-250$ Hz) as the $B$-field is tuned from 10 to 600 G. The variations of  $J_\perp$ observed in Fig.~4 match those in the transition EDMs shown in Figs.~1 and 3.
 In particular, the avoided crossing between the eigenstates $|10\rangle$ and $|11\rangle$ causes the spin-exchange couplings $J_\perp^{3,10}$ and $J_\perp^{3,11}$ to switch at $B=140$ G. 
 Similar sensitivity of $J_\perp$ to the $B$ field is observed for the eigenstates in the $M_F=-3/2$ symmetry sector, as shown in Fig. 4(b), and in the presence of an $E$ field [see Figs.~4(c) and (d)].

The strong dependence and fast change of $J_{\perp}$ with magnetic field opens a unique opportunity to use molecules for d.c. magnetometry. This can be achieved by taking  advantage of the so called density shift or dipolar induced precession of the collective Bloch vector of the system in the presence of ED interactions \cite{Chang:04,Li:23}. The latter  can be measured in a standard Ramsey spectroscopy sequence, where $\mathcal{N}$ molecules prepared in the lowest rotational state are illuminated by a microwave drive to generate a coherent superposition with a targeted  excited rotational state  (denoted as   $|\uparrow\rangle$ state) i.e. 
$|\psi(0)\rangle= (\cos(\Theta)|\downarrow\rangle+\sin(\Theta)|\uparrow\rangle)^\mathcal{N}$. In this case the $|\uparrow\rangle$ state is the one that features the sharp resonance   as a function of $B$.  After the pulse,  the system is let to evolve under the presence of dipolar exchange interactions  for time $t$, accumulating a density shift, which manifests by the accumulated phase $ \phi(t)= (J_z-J_\perp ) \eta t \cos(\Theta)$
and $ \eta = \sum_{j} (1-3\cos^2\theta_{0j})/|{\bf R}_{0j} |^3 $. The phase is measured via  a second $\pi/2$ pulse that  converts it  into a  population difference.  As any standard Ramsey sequence, the sensitivity of this protocol   for an array of $N$ independent molecules is given by the so called standard quantum limit $ \Delta \phi (t) =1/\sqrt{\mathcal{N}}$.
The fact that $\phi$ is a very sensitive function of $B$ nevertheless opens a path for very precise magnetic field sensing with a sensitivity given by $\Delta B=  \Big( \frac { \sqrt{\mathcal{N}} d   \phi}{dB}\Big)^{-1}$ with $ \frac {  d   \phi}{dB}= \eta t\frac{ d J_\perp}{dB }$. For unit filled molecular arrays in 2D geometries with the electric field perpendicular to the molecule plane, the  optimal  sensitivity scales as $\Delta B=  \frac {1} { \cos (\Theta) t \eta \sqrt{\mathcal{N}}} \frac{dB}{dJ_\perp}$  which can be as large as $\Delta B= 1/( 2\eta \cos\Theta \sqrt{\mathcal{N}}$) $ \mu$T at one second close to the point of maximum slope (around $B=152.2$~G, $\frac{dB}{dJ_\perp}\sim 200 $G/Hz). 
This  translates into a sensitivity at the level  of a few hundred pT /$\sqrt{\rm Hz}$  for an array of pinned $10^5$ molecules assuming $\eta \sim 1$.  Even though this  value is at least three orders of magnitude less sensitive than that achievable with state of the art cold atom magnetometers \cite{Bai2023}, it potentially  offers unique opportunities for improvement given the many-body nature of the shift. For example, by enhancing the range of the  exchange interactions via the use of microwave cavities  or by operating with itinerant arrays instead of pinned particles, $\eta$ could  be made to scale linearly  with $\mathcal{N}$, significantly increasing the achievable  sensitivity. 

Finally, our results suggest novel possibilities for high-dimensional quantum information processing \cite{Wang:20,Sawant:20} and quantum simulation \cite{GonzalezCuadra:22} with ultracold polar molecules. 
As shown in Figs.~1 and 3, the splitting between the $N=1$ levels near an avoided crossing can be made smaller than $\simeq50$~Hz, the strength of the ED interaction at a typical lattice spacing. As a result, molecules near such crossings can no longer be described as effective two-level systems, and the explicit inclusion of the third level becomes necessary. To this end, we describe each molecule by an effective three-level  system (qutrit) comprising the ground ($\ket{0}$) and two excited ($\ket{-1})$ and $\ket{1}$) energy levels in the V-configuration. The effective ED interaction between the  three-level systems at the avoided crossing, where the transition EDMs are both equal to $d$ [see Fig. 1(a)], takes the form 
\begin{equation}\label{eq:H3LS}
\hat{H}_{ij}= 
\frac{d^2(1 - 3\cos^2\theta_{ij})}{R_{ij}^3}
\left[\hat{S}_x^i \hat{S}_x^j+\hat{Q}_{yz}^i \hat{Q}_{yz}^j\right],
\end{equation}
where $\hat{S}^i_\alpha$ are the effective dipole (or spin-1) operators, and $\hat{Q}^i_{\alpha\beta}$ ($\alpha,\beta=x,y,z$) are the quadrupole operators, which form an orthogonal basis of the {\it su}(3) Lie algebra \cite{Di:10,Hamley:12}, whose elements are infinitesimal generators of the SU(3) Lie group of unitary single-qutrit gates \cite{LeBlanc:23}.
The Hamiltonian \eqref{eq:H3LS} contains similar processes to the ones engineered with effective all-to-all couplings  in spinor quantum gases of ultracold atoms (such as four-wave mixing) \cite{Hamley:12}, where spin-nematic squeezed vacuum has been experimentally realized thanks to the SU(2)-like character of the $\{\hat{S}_x,\hat{Q}_{yz},\hat{Q}_{xz} \}$ quadratures \cite{Bilitewski:23}. The implementation of this Hamiltonian in dipolar molecules can open unique opportunities of realizing such states with even richer properties.



In summary, we have shown that transition EDMs of polar molecules  can be magnetically tuned over a wide dynamic range, and can even be made to vanish as shown in Fig.~1(a) and (d), effectively
turning a polar molecule like KRb into a non-polar one!
The underlying mechanism relies on  
 narrow avoided crossings and the quantum ergodic behavior of molecular eigenstates mediated by the NEQ interaction.
This enables continuous magnetic tuning of exchange ED interactions between zero and a maximum value without the need to transfer the molecules from one quantum state to another. 
Our approach requires neither strong magnetic fields nor microwave dressing, and only relies on the interplay between the hyperfine and Zeeman interactions. As a result, it can be applied to 
 laser-coolable $^2\Sigma$ molecules \cite{Barry:14,Collopy:18,McCarron:18,Changala:19,Anderegg:18,Burau:23} and even polyatomic molecules \cite{Prehn:16,Kozyryev:17,Mitra:20,Augenbraun:20,Vilas:22,Liu:23,Zhang:23} providing  a versatile tool for controlling intermolecular interactions in the quantum regime.


We are grateful for stimulating discussions with  Jun Ye, James Thompson,  and Nathan Prins  and for feedback on the manuscript by  Junyu Lin and  Lee Liu.
This work was supported by the NSF EPSCoR RII Track-4 Fellowship No. 1929190, the AFOSR MURI, the NSF JILA-PFC PHY-2317149, and by NIST.

\newpage

\widetext
\begin{center}
\textbf{\large Supplemental Material}
\end{center}
\setcounter{equation}{0}
\setcounter{figure}{0}
\setcounter{table}{0}
\title{ \bf{Supplemental Material for \\  ``Magnetically tunable electric dipolar interactions of ultracold polar \\molecules in the quantum ergodic regime''}}

\author{ Rebekah Hermsmeier$^{1}$, Ana Maria Rey$^{2}$, and Timur V. Tscherbul$^{1}$ \\ }

\vspace{0.5cm}
\affiliation{ \vspace{0.3cm}
$$^{1}$Department of Physics, University of Nevada, Reno, Nevada, 89557, USA\\
$^{2}$JILA, National Institute of Standards and Technology, and Department of Physics, University of Colorado,  Boulder, Colorado, 80309, USA}
\maketitle

\tableofcontents

\vspace{0.8cm}

In this Supplemental Material, we present additional calculations of the magnetic field dependence of the EDMs for a range of rotational transitions in KRb in the quantum ergodic regime (Sec.~I). These calculations support the general claim made in the main text that the EDMs   can be effectively controlled by an external magnetic field.
In Sec.~II we discuss the matrix elements of the EDM operators $\hat{d}_+$ and $\hat{d}_-$.
 Section III  discusses the effect of high electric fields on ergodicity. Finally, Sections IV and V provide a detailed derivation of the effective $S = 1$  Hamiltonian for two three-level molecules near an avoided crossing of two $N=1$ energy levels (see Figs.~1 and 2 of the main text) interacting via the electric dipolar interaction.

\section{Additional Examples of magnetic tunability of transition EDMs}
\begin{figure}[]
 \centering
 \includegraphics[width=8.5cm]{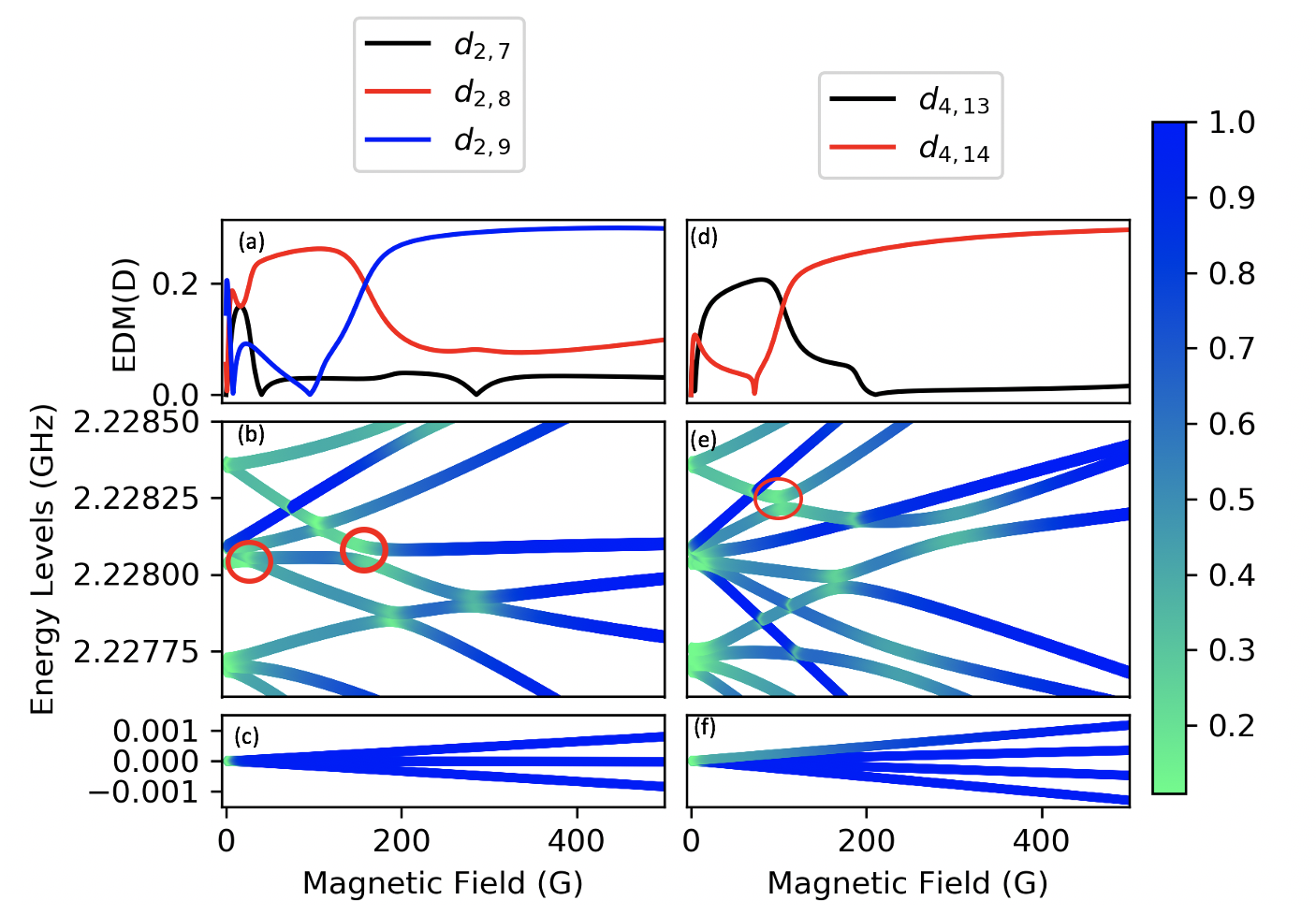}
 \caption{ Representative transition EDMs $\langle x|d|y\rangle$ and $\langle x_1|d|y_1\rangle$ plotted as a function of magnetic field at zero electric field for {$M_F= -7/2$ (a) and $M_F=-1/2$ }(d). The corresponding $N=0$ and $N=1$ hyperfine-Zeeman energy levels are shown in panels (b), (c), (d) and (e), respectively. The color of the energy levels corresponds to 1 minus the maximum inverse participation ratio of the corresponding energy eigenstates (see the main text). Encircled are the avoided crossings responsible for the   swapping of the corresponding transition EDMs. }
  \label{fig:EDMpart2}
\end{figure}

Here, we provide additional calculations of transition EDMs for KRb molecules to illustrate that their wide magnetic tunability is common. Figures~1 (a) and (d) show the transition EDMs of KRb as a function of magnetic field at zero electric field. In these figures, both $\langle 2|d_0|9\rangle$ and $\langle 4|d_0|14\rangle$ exhibit a rapid increase from zero to a nearly constant value as the magnetic field increases from 96 to 210 G and from 73 to 160 G, respectively. Notably, sharp variations in transition dipoles are often accompanied by avoided crossings occurring at the same magnetic field.

Figure 1(b) highlights an avoided crossing between energy eigenstates $|8\rangle$ and $|9\rangle$ at the magnetic field, where a significant increase in the EDM matrix element $\langle 2|d_0|9\rangle$ and the corresponding decrease in $\langle 2|d_0|8\rangle$ occur. A similar behavior is observed in the transition EDMs $\langle 4|d_0|13\rangle$ and $\langle 4|d_0|14\rangle$ in Figure 1(e). These avoided crossings, akin to those discussed in the main text, result from the nuclear electric quadrupole (NEQ) interaction. Additionally, they induce substantial changes in the characteristics of the involved eigenstates, a phenomenon explained by the model presented in the main text. We find that such occurrences are so prevalent that, for every  total angular momentum projection $M_F$ and every $N=0$ sublevel, there exists at least one transition EDM to the  corresponding sublevel in the $N=1$ manifold, which can be tuned from zero to its maximum value.

Figure 2 shows the calculated transition EDMs for $M_F=-9/2$ (a) and $M_F=-5/2$ (d). In panel (a),  the matrix element $\langle 2|d_0|7\rangle$ smoothly increases from 0 to 0.3 D as the magnetic field varies from 227 to 1000 G. In panel (d), the EDM $\langle 3|d_0|11\rangle$ is larger at low magnetic fields, decreasing at 87 G, where $\langle 3|d_0|11\rangle$ starts to increase. We further observe that at $B=412$~G, $\langle 3|d_0|11\rangle$ starts to decrease again as $\langle 3|d_0|12\rangle$ increases. These changes in the transition EDMs occur at the same $B$ fields as the avoided crossings in panels (b) and (e), further demonstrating the extensive range of conditions, where significant magnetic tunability of EDMs is achievable.

\begin{figure}[]
 \centering
 \includegraphics[width=8.5cm]{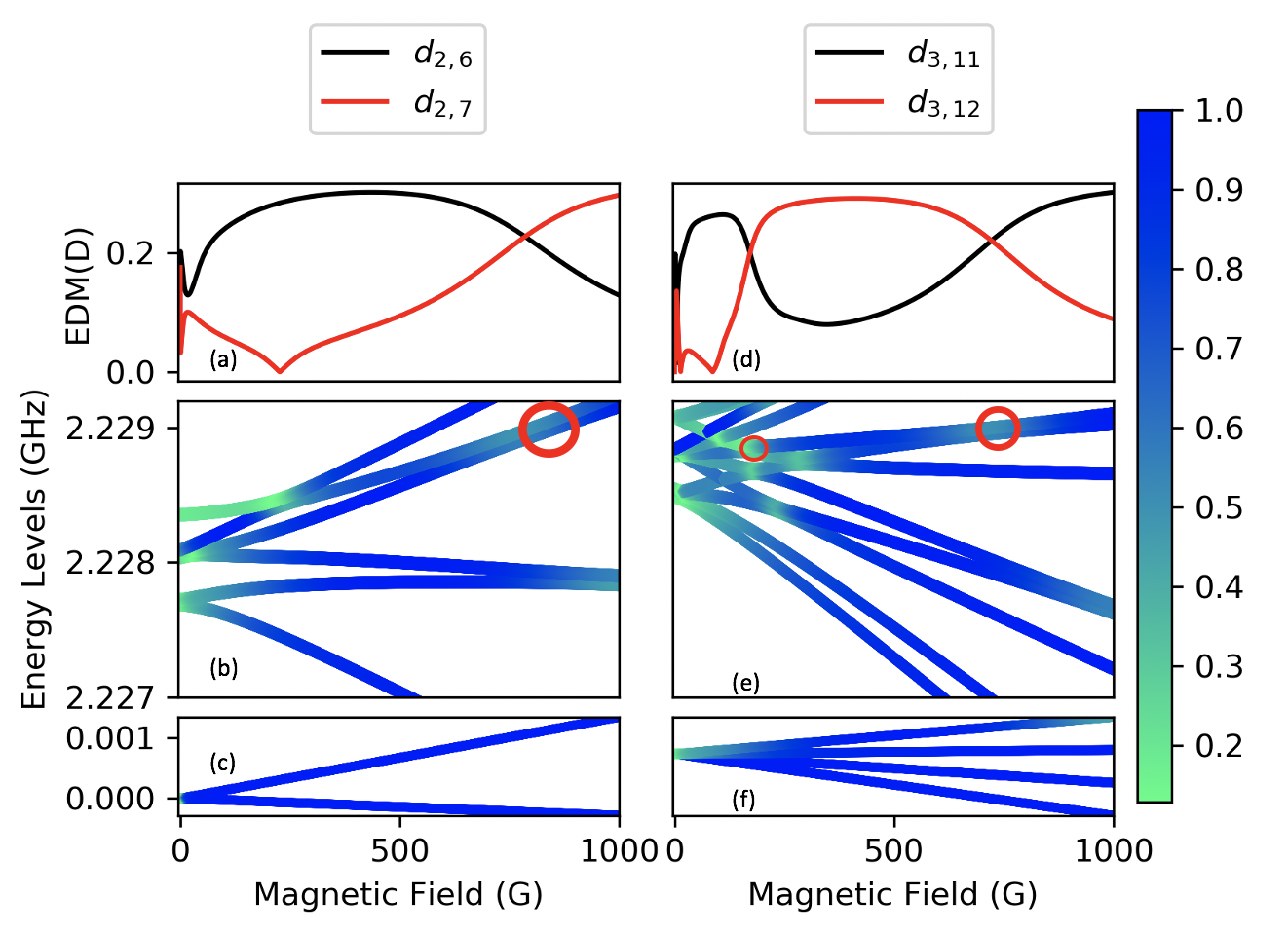}
 \caption{Transition EDMs plotted as a function of magnetic field at zero electric field for $M_F= -9/2$ (a) and $M_F=-5/2$ (d). The corresponding $N=0$ and $N=1$ hyperfine-Zeeman energy levels are shown in panels (b), (c), (d) and (e), respectively. The color of the energy levels corresponds to the ergodicity of the energy eigenstates (see the main text), where green indicates regions of high ergodicity. The avoided crossings responsible for the changes in transition EDMs are encircled. 
 }
  \label{fig:EDMpart3}
\end{figure}

\begin{figure}[]
 \centering
 \includegraphics[width=8.5cm]{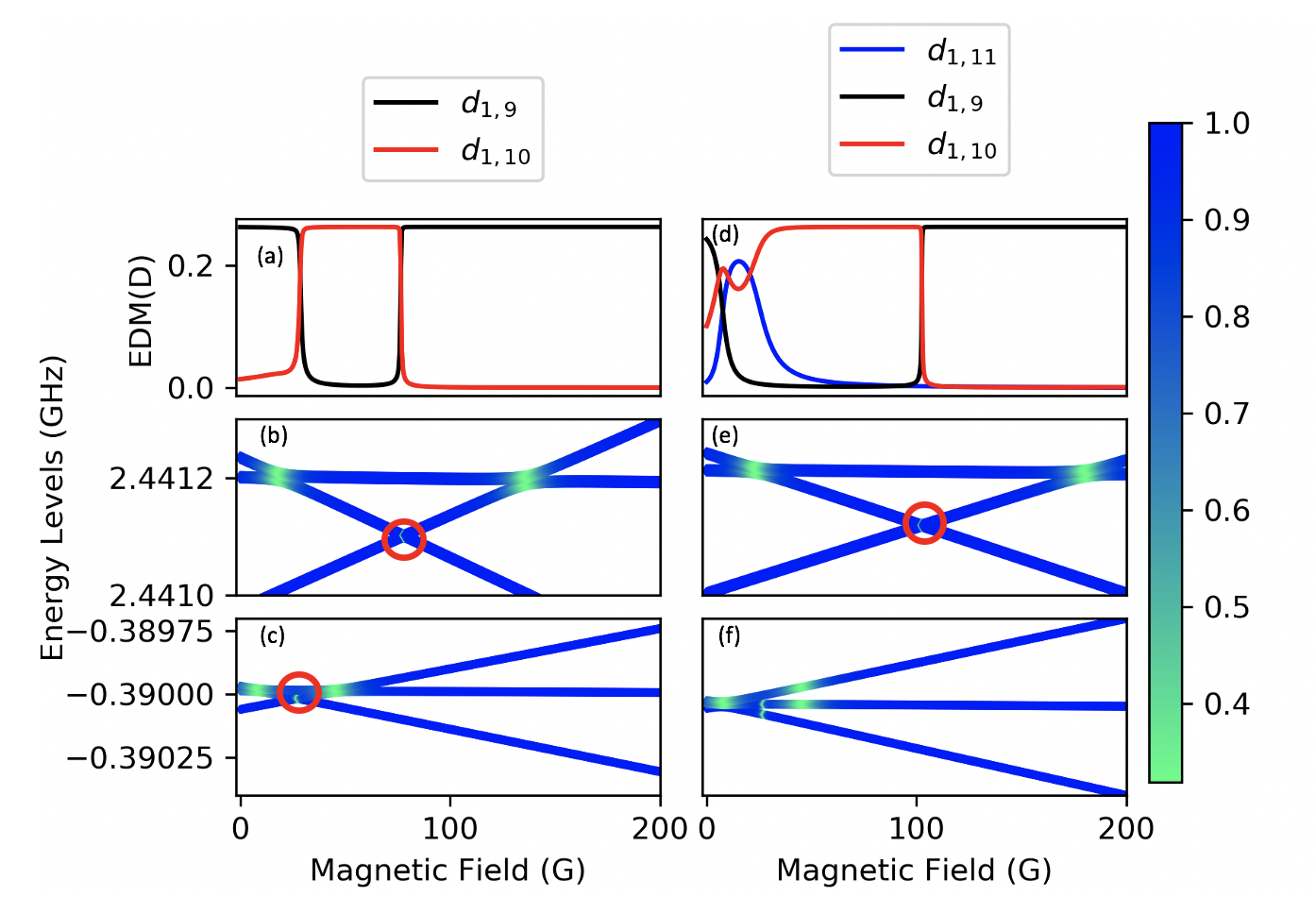}
 \caption{Transition EDMs $\langle 1|d|k\rangle$ ($k=9-11$) plotted as a function of magnetic field for $M_F= -7/2$ at electric fields of 6 kV/cm (a) and 2 kV/cm (d). The corresponding $N=0$ hyperfine-Zeeman energy levels are shown in panels (c) and (e), respectively. The relevant $N=1$ levels (e.g. 9-11) are shown in panels (b) and (d). The color gradient shows the ergodicity of the energy levels, where green indicates that energy levels are highly ergodic. The red circles indicate the avoided crossings between the $N=0$  and $N=1$ states responsible for changes in the transition EDMs.  
}
  \label{fig:EDMpart4}
\end{figure}


Figure 3 shows the transition EDMs of KRb as a function of magnetic field at electric fields of 6 kV/cm (a) and 2~kV/cm (d). In panel (d), the transition EDM $\langle 1|d_0|10\rangle$ exhibits a sharp decrease from the maximum value to zero at 120 G. In comparison, in panel (a), $\langle 1|d_0|10\rangle$ experiences two sharp variations at $E=2$~kV/cm. 
The initial sharp change in the EDM corresponds to an avoided crossing between two $N=0$ energy states, a feature absent at lower electric fields [see e.g., panel (d). Therefore, at high electric fields, magnetic tunability becomes possible through two mechanisms: avoided crossings between the Stark states in the $N=0$  manifold, and between those  in the $N=1$ manifold. 
We note that in the presence of a dc $E$ field, $N$ is not a rigorously good quantum number. However, since here we are interested in relatively weak $E$ fields such that $Ed/B_e < 1$, $N$ remains an approximately good quantum number in the sense that molecular Stark states  have well-defined values of $N$ in the limit $E\to 0$.

\section{Magnetic tuning of the $\hat{d}_\pm$ spherical components}

Thus far, our discussion has focused on the matrix elements of the EDM operator $\hat{d}_0$ because 
the matrix elements of the other spherical tensor components of $\hat{\mathbf{d}}$ ($\hat{d}_\pm$) are zero between the initial and final eigenstates, for which the matrix elements of $\hat{d}_0$ are nonzero \cite{Asnaashari:23} owing to the conservation of $M_F$.
Figures 4 (a) and (d) show the off-diagonal matrix elements of $\hat{d}_\pm$ as a function of magnetic field at zero electric field. The transition EDMs $\langle 8|\hat{d}_+|1\rangle$  and $\langle 7|\hat{d}_+|1\rangle$ vary smoothly, interchanging values near $B=200$~G.
This interchange  is caused by the avoided crossing between the $N=1$ states $|7\rangle$ and $|8\rangle$, which alters the character of the energy eigenstates. In panel (d), $\langle 8|\hat{d}_-|1\rangle$ and $\langle 11|\hat{d}_-|1\rangle$ experiences sharp changes in value at 80 and 175 G, respectively. Additionally, the matrix elements $\langle 9|\hat{d}_-|1\rangle$  and $\langle 10|\hat{d}_-|1\rangle$ undergo two sharp changes at 80 and 106 G, and at 106 and 175 G, respectively. Each of these changes corresponds to an avoided crossing. Thus, the presence of avoided crossings  between the $N=1$ states allows for magnetic tunability of the matrix elements of $\hat{d}_+$ and $\hat{d}_-$, just as in the case of the matrix elements of  $\hat{d}_0$ discussed in the main text. 


\begin{figure}[]
 \centering
 \includegraphics[width=8.5cm]{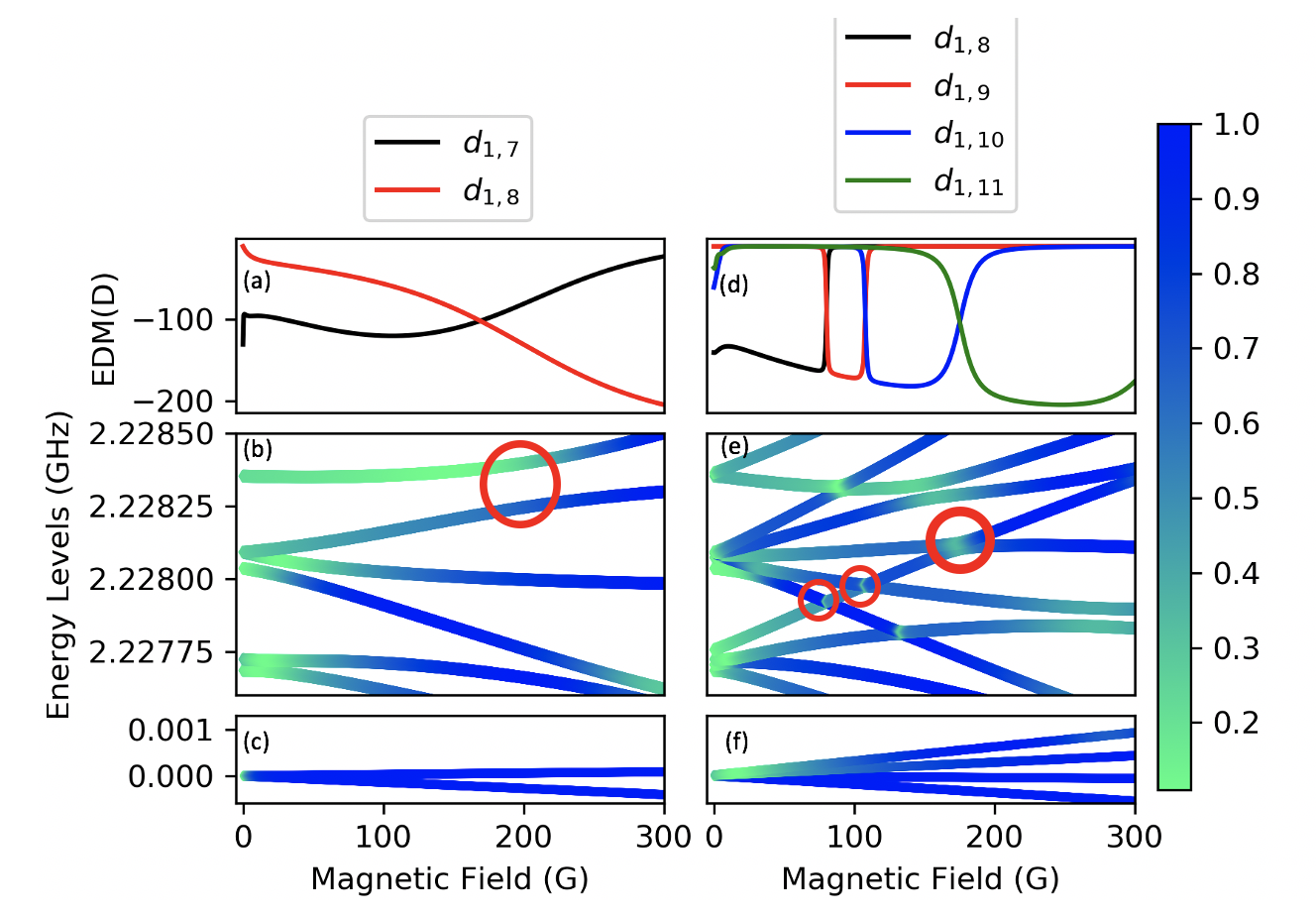}
 \caption{Transition EDMs {$\langle x (M_F=9/2)|d_+|y(M_F=7/2)\rangle$ }(a)and {$\langle x(M_F=5/2)|d_-|y(M_F=7/2)\rangle$} (d) are plotted as a function of magnetic field at 0 kV/cm electric field. The corresponding $N=0$ and $N=1$ hyperfine-Zeeman energy levels are shown in panel (b), (c), (d) and (e), respectively.  Encircled are the avoided crossings between $N=1$ levels responsible for the switching behavior of the EDMs in panels (a) and (c).
}
  \label{fig:Jperp4}
\end{figure}

\section{Ergodicity of KRb eigenstates at high electric fields}


Figures 5 (a) and (e) compare the magnetic field dependence of transition EDMs calculated in low vs. high dc electric fields.
We observe that the energy levels at 12 kV/cm are more ergodic  than those at 1 kV/cm. This occurs because the Stark effect couples the $N=0$ ground states with the $N=1,M_N=0$ excited states, leading to an increasingly mixed character of the eigenstates  at higher $E$ fields. Interestingly, this effect is more pronounced for the states in the $N=0$ manifold, where a significant spike in ergodicity is observed near $B=100$~G. This is due to a combination of two effects. First, there is an avoided crossing between two $N=1$ excited states induced by the NEQ interaction. Second, the Stark effect couples the ground $N=0$ states with one of the excited $N=1$ states involved in the avoided crossing, causing an increase in ergodicity.

\begin{figure}[]
 \centering
 \includegraphics[width=15.5cm]{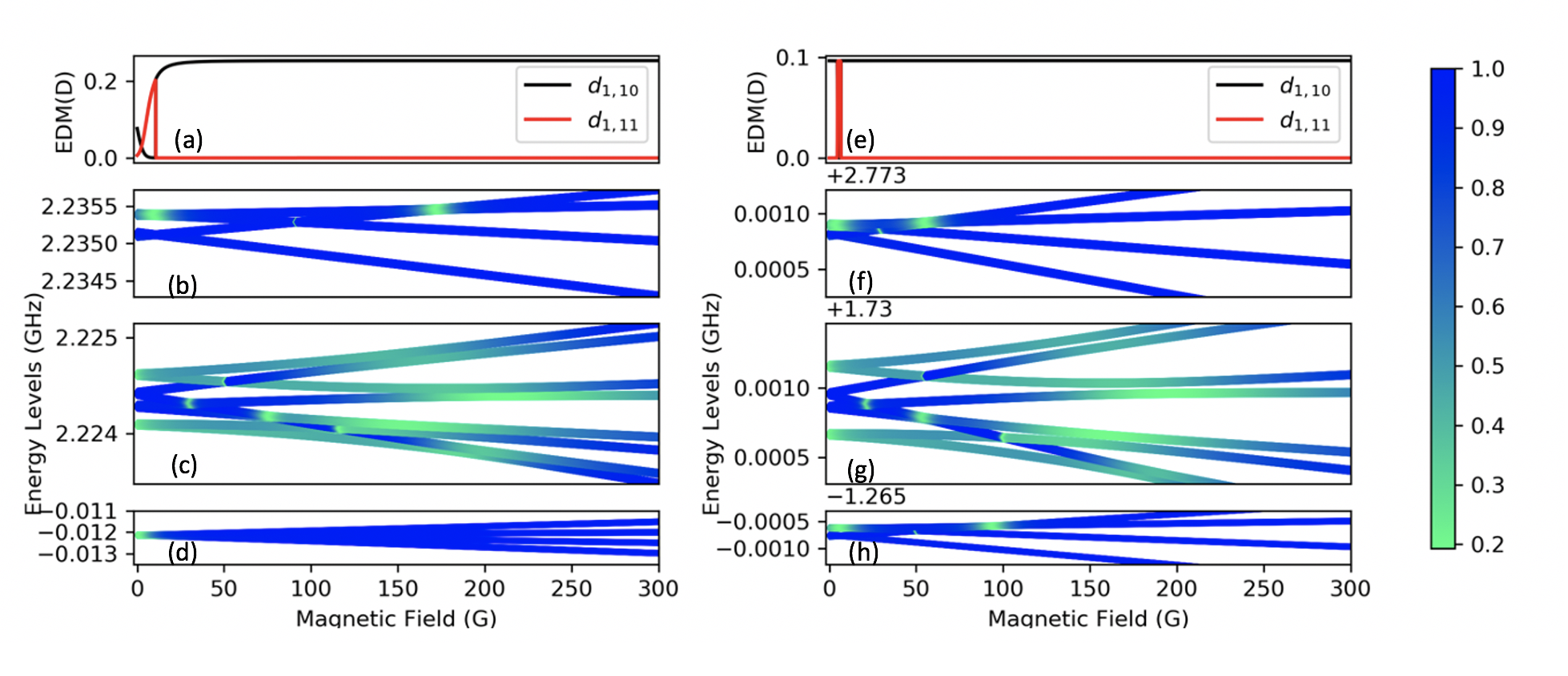}
 \caption{Transition EDMs $\langle 1|d|10\rangle$ and $\langle 1|d|11\rangle$ plotted as a function of magnetic field for $M_F= -3/2$ at electric fields of 1 kV/cm (a) and 12 kV/cm {(e)}. The hyperfine-Zeeman energy levels for $N=0$ are illustrated in panels (d) and (h), while the $N=1$ hyperfine-Zeeman energy levels are presented in panels {(b), (c), (f), and (g)}, with panels (b) and {(f)} specifically depicting the energy levels corresponding to $M_N=0$.
 }
  \label{fig:Jperp5}
\end{figure}

\section{Derivation of the effective S = 1 Hamiltonian}

In this section, we derive an effective spin-lattice Hamiltonian for a system of three-level molecules trapped in an optical lattice and interacting via the electric dipole-dipole (ED) interaction. Our effective three-level system is formed by the $N=0$ eigenstate $|3\rangle$ and two $N=1$ eigenstates $|10\rangle$ and $|11\rangle$,  which are nearly degenerate at $B=140$~G, forming an avoided crossing shown in Fig.~1(b) of the main text. The transition EDMs between the nearly degenerate $N=1$ states and state $|3\rangle$ are equal at the crossing, as illustrated in Fig.~1(a) of the main text,  creating an effective three-level V-system. Note that there is no direct coupling between the rotationally excited states (see Fig.~6).


\begin{figure}[]
 \centering
 \includegraphics[width=8.5cm]{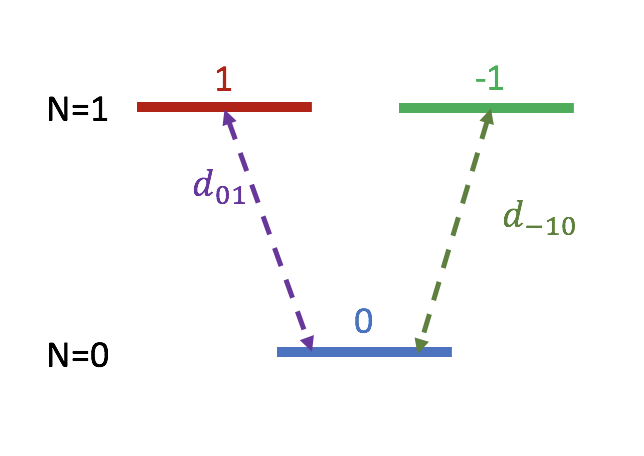}
 \caption{The effective three-level system in the V configuration. The ground state $|0\rangle$ is coupled by the transition EDMs $d_{01}$ and $d_{-10}$ to the nearly degenerate states $|1\rangle$ and $|-1\rangle$, which form an avoided crossing.}
  \label{fig:Jperp7}
\end{figure}



The effective Hamiltonian of our three-level molecule may be written as $\hat{H}_\text{mol}=\sum_{n=-1,0,1}E_n |n\rangle \langle n|$, where $E_n$ are the eigenenergies shown in Fig.~6, and $|n\rangle$ are the corresponding eigenstates.
Projecting the dipole moment operator of the $i$-th molecule onto these eigenstates, we obtain (assuming  $E=0$, so that $d_{nn}=0)$
\begin{equation}
\begin{aligned} 
\hat{d}_0^{(i)}=\sum_{m=1,0,-1}\sum_{n=1,0,-1}d_{mn}^{(i)} |m_i\rangle \langle n_i|\\
=d_{10}^{(i)}|1_i\rangle \langle 0_i|+d_{-10}^{(i)} |-1_i\rangle \langle 0_i|+d_{0-1}^{(i)} |0_i\rangle \langle -1_i|+d_{01}^{(i)} |0_i\rangle \langle 1_i|
\end{aligned}
\end{equation}
Using this expansion, we can express the $q=0$ part of the ED interaction between the $i$-th and $j$-th molecules
\begin{equation}\label{eq:Hdd1}
\begin{aligned} 
\frac{1- 3\cos^2\theta_{ij}}{R_{ij}^3}\bigg(\hat{d}_0^{(i)}\otimes \hat{d}_0^{(j)}\bigg)=\frac{1- 3\cos^2\theta_{ij}}{R_{ij}^3}\Big(d_{10}^{(i)}|1_i\rangle \langle 0_i|+d_{-10}^{(i)} |-1_i\rangle \langle 0_i|+d_{0-1}^{(i)} |0_i\rangle \langle -1_i|+d_{01}^{(i)} |0_i\rangle \langle 1_i|)\otimes\\ (d_{10}^{(j)}|1_j\rangle \langle 0_j|+d_{-10}^{(j)} |-1_j\rangle \langle 0_j|+d_{0-1}^{(j)} |0_j\rangle \langle -1_j|+d_{01}^{(j)} |0_j\rangle \langle 1_j|\Big),
\end{aligned}
\end{equation}
where $R_{ij}$ is the distance between the molecules and  the angle $\theta_{ij}$ is  defined in the main text.
At the avoided crossing, the transition EDMs of our V-system are equal, $d_{01}=d_{-10}=d_t$. Expanding Eq.~\eqref{eq:Hdd1} we obtain
\begin{equation}\label{eq:Hdd2}
\begin{aligned} 
\frac{1- 3\cos^2\theta_{ij}}{R_{ij}^3}\bigg(\hat{d}_0^{(i)}\otimes \hat{d}_0^{(j)}\bigg)=\frac{1- 3\cos^2\theta_{ij}}{R_{ij}^3} d_t^2\bigg( \textcolor{red}{ |1_i\rangle \langle 0_i| \otimes|1_j\rangle \langle 0_j|}
+\textcolor{red}{ |-1_i\rangle \langle 0_i|\otimes |1_j\rangle \langle 0_j|}+|0_i\rangle \langle -1_i|\otimes |1_j\rangle \langle 0_j|\\
+ |0_i\rangle \langle 1_i|\otimes  |1_j\rangle \langle 0_j|
+\textcolor{red}{|1_i\rangle \langle 0_i|\otimes |-1_j\rangle \langle 0_j|}+\textcolor{red}{ |-1_i\rangle \langle 0_i|\otimes |-1_j\rangle \langle 0_j|}\\
+ |0_i\rangle \langle -1_i|\otimes |-1_j\rangle \langle 0_j|+ |0_i\rangle \langle 1_i|\otimes |-1_j\rangle \langle 0_j|
+|1_i\rangle \langle 0_i|\otimes |0_j\rangle \langle -1_j|\\
+ |-1_i\rangle \langle 0_i|\otimes |0_j\rangle \langle -1_j|
+\textcolor{red}{ |0_i\rangle \langle -1_i|\otimes|0_j\rangle \langle -1_j|}+\textcolor{red}{|0_i\rangle \langle 1_i|\otimes |0_j\rangle \langle -1_j|}\\
+|1_i\rangle \langle 0_i|\otimes |0_j\rangle \langle 1_j|+ |-1_i\rangle \langle 0_i| \otimes |0_j\rangle \langle 1_j|
+\textcolor{red}{|0_i\rangle \langle -1_i|\otimes |0_j\rangle \langle 1_j|}\\
+\textcolor{red}{ |0_i\rangle \langle 1_i|\otimes |0_j\rangle \langle 1_j|}\bigg)
\end{aligned}
\end{equation}
The  terms highlighted in red indicate the processes, which do not conserve energy. For instance, the first such term $ \textcolor{red}{ |1_i\rangle \langle 0_i| \otimes|1_j\rangle \langle 0_j|}$  transfers  the molecules $i$ and $j$ from the state $|0_i\rangle|0_j\rangle$  to the state $|1_i\rangle|1_j\rangle$, a process that does  not conserve energy.  Neglecting these energetically off-resonant processes,  Eq.~\eqref{eq:Hdd2} becomes
 \begin{equation}
\begin{aligned} 
\frac{1- 3\cos^2\theta_{ij}}{R_{ij}^3}\bigg(\hat{d}_0^{(i)}\otimes \hat{d}_0^{(j)}\bigg)=\frac{1- 3\cos^2\theta_{ij}}{R_{ij}^3} d_t^2\bigg( 
|0_i\rangle \langle -1_i|\otimes |1_j\rangle \langle 0_j|
+ |0_i\rangle \langle 1_i|\otimes  |1_j\rangle \langle 0_j|
+ |0_i\rangle \langle -1_i|\otimes |-1_j\rangle \langle 0_j|\\
+ |0_i\rangle \langle 1_i|\otimes |-1_j\rangle \langle 0_j|
+|1_i\rangle \langle 0_i|\otimes |0_j\rangle \langle -1_j|
+ |-1_i\rangle \langle 0_i|\otimes |0_j\rangle \langle -1_j|\\
+|1_i\rangle \langle 0_i|\otimes |0_j\rangle \langle 1_j|+ |-1_i\rangle \langle 0_i| \otimes |0_j\rangle \langle 1_j|
\bigg)
\end{aligned}
\end{equation}
Combining like terms, we find
\begin{equation}
\begin{aligned} 
\frac{1- 3\cos^2\theta_{ij}}{R_{ij}^3}\bigg(\hat{d}_0^{(i)}\otimes \hat{d}_0^{(j)}\bigg)
=\frac{1- 3\cos^2\theta_{ij}}{R_{ij}^3} d^2_t\bigg( \big(|1_i\rangle \langle0_i|+|-1_i\rangle \langle0_i|\big)\otimes \big(|0_j\rangle \langle-1_j|+|0_j\rangle \langle1_j|\big)\\
+\big(|0_i\rangle \langle-1_i|+|0_i\rangle \langle1_i|\big)\otimes \big(|1_j\rangle \langle0_j|+|-1_j\rangle \langle0_j|\big)\bigg).
\end{aligned}
\end{equation}

We now wish to map the operators that occur in the above equation onto the effective spin operators acting in the Hilbert subspaces of the $i$-th and $j$-th molecules. The effective spin operators for the three-level system can be expressed in terms of the generators of the SU(3) Lie group, which are commonly chosen as either the Gell-Mann matrices  \cite{ZeeBook}, or, equivalently, as linear and biquadratic functions of spin-1 operators  $\hat{S}_i$ ($i = x, y, z$) \cite{Richarsdon:20}.

We choose the latter representation to make contact with the theory of spinor Bose-Einstein condensates \cite{Chapman:12,PethickBook}. To this end, we define the standard spin-1 operators $\hat{S}_i$ ($i = x, y, z$) along with the nematic moments $\hat{Q}_{ij}$, which are related to the spin-1 operators as $\hat{Q}_{ij}=\hat{S}_i\hat{S}_j + \hat{S}_j \hat{S}_i-\frac{4}{3}\delta_{ij}$.


All of the above operators can be expressed in terms of the jump operators $|m\rangle\langle n|$:
\begin{equation}
\begin{aligned} 
\hat{S}_x=\frac{1}{\sqrt{2}}\bigg(|1\rangle \langle 0|+|0 \rangle \langle -1|+|0 \rangle \langle 1|+|-1\rangle \langle 0|\bigg)\\
\hat{S}_y=\frac{i}{\sqrt{2}}\bigg(-|1\rangle \langle 0|-|0 \rangle \langle -1|+|0\rangle \langle 1|+|-1\rangle \langle 0|\bigg)\\
\hat{S}_z=|1\rangle \langle 1|-|-1\rangle \langle -1|\\
\hat{S}_+=\sqrt{2}\bigg(|1\rangle \langle 0|+|0 \rangle \langle -1|\bigg)\\
\hat{S}_-=\sqrt{2}\bigg(|0\rangle \langle 1|+|-1\rangle \langle 0|\bigg)\\
\hat{Q}_{xz}=\frac{1}{\sqrt{2}} \bigg(|1 \rangle \langle 0|-|0 \rangle \langle -1|+|0 \rangle \langle 1|-|-1\rangle \langle 0|\bigg)\\
\hat{Q}_{xy}=i \bigg(-|1 \rangle \langle -1|+|-1\rangle \langle 1|\bigg)\\
\hat{Q}_{yz}=\frac{i}{\sqrt{2}} \bigg(-|1\rangle \langle 0|+|0 \rangle \langle -1|+|0 \rangle \langle 1|-|-1\rangle \langle 0|\bigg)
\end{aligned}
\end{equation}

The ED interaction  can now be recast as the effective spin-spin interaction involving the spin-1 operators of the $i$-th and $j$-th molecule
\begin{equation}
\begin{aligned} 
\frac{1- 3\cos^2\theta_{ij}}{R_{ij}^3}\bigg(\hat{d}_0^{(i)}\otimes \hat{d}_0^{(j)}\bigg)
=\frac{1- 3\cos^2\theta_{ij}}{R_{ij}^3} d^2_t\bigg( \big(|1_i\rangle \langle0_i|+|-1_i\rangle \langle0_i|\big)\otimes \big(|0_j\rangle \langle-1_j|+|0_j\rangle \langle1_j|\big)\\
+\big(|0_i\rangle \langle-1_i|+|0_i\rangle \langle1_i|\big)\otimes \big(|1_j\rangle \langle0_j|+|-1_j\rangle \langle0_j|\big)\bigg)\\
=d^2_t \frac{1- 3\cos^2\theta_{ij}}{R_{ij}^3}\bigg( \frac{1}{\sqrt{2}}(\hat{S}_x^i+i\hat{Q}_{yz}^i)\otimes \frac{1}{\sqrt{2}}(\hat{S}_x^j-i\hat{Q}_{yz}^j)+\frac{1}{\sqrt{2}}(\hat{S}_x^i-i\hat{Q}_{yz}^i)\otimes \frac{1}{\sqrt{2}}(\hat{S}_x^j+i\hat{Q}_{yz}^j)\bigg)\\
=\frac{d^2_t}{2}\frac{1- 3\cos^2\theta_{ij}}{R_{ij}^3}\bigg( (\hat{S}_x^i+i\hat{Q}_{yz}^i)\otimes (\hat{S}_x^j-i\hat{Q}_{yz}^j)+(\hat{S}_x^i-i\hat{Q}_{yz}^i)\otimes(\hat{S}_x^j+i\hat{Q}_{yz}^j)\bigg)\\
=d_t^2 \frac{1- 3\cos^2\theta_{ij}}{R_{ij}^3} \bigg(\hat{S}_x^i \otimes \hat{S}_x^j+\hat{Q}_{yz}^i\otimes \hat{Q}_{yz}^j\bigg)..
\end{aligned}
\end{equation}
In Section V, we employ an alternative method to evaluate this Hamiltonian.

\section{Derivation of the effective $S = 1$ Hamiltonian using a matrix approach}

To verify our findings in the previous section, here we  provide an alternative derivation of the effective spin-spin Hamiltonian for two three-level V-type molecules interacting via the ED interaction.  As before, our starting point is the general Hamiltonian for the ED interaction
\begin{equation}\label{eq:Hdd3terms}
\begin{aligned} 
\hat{H}_{DD}=\frac{1-3\cos^2\theta_{ij}}{R_{ij}^3}(\hat{d}_0^{(i)}\hat{d}_0^{(j)}+\frac{d_1^{(i)}d_{-1}^{(j)}+d_{-1}^{(i)}d_1^{(j)}}{2})
\end{aligned}
\end{equation}
We note that the second and third terms involving the products of the $\hat{d}_{\pm1}$ operators can be neglected for our $V$-system configuration because it requires an avoided crossing between the excited energy levels.
In order for two states to experience an avoided crossing, they must have the same projection of total angular momentum $M_F$. Furthermore, it is essential to recognize that EDM cannot couple states with differing nuclear spin projections. For example, consider the EDM operator $\hat{d}_1$, which can couple the states $|00,4,-1/2\rangle$ and $|11,4,-1/2\rangle$, while $d_{-1}$ can couple the states $|00,4,-1/2\rangle$ and $|1-1,4,-1/2\rangle$. However, due to the incompatible spin projections, the states $|1-1,4,-1/2\rangle$ and $|11,4,-1/2\rangle$ cannot undergo an avoided crossing. Consequently, the terms $\hat{d}_1 \hat{d}_{-1}$ and $\hat{d}_{-1} \hat{d}_1$ do not couple the states of our V-system, and the Hamiltonian \eqref{eq:Hdd3terms} can be simplified to $H_{DD}=\frac{1-3\cos^2\theta_{ij}}{r_{ij}^3}\hat{d}_0^{(i)}\hat{d}_0^{(j)}$, allowing us to calculate its matrix elements.
\begin{equation}
\begin{aligned} 
\langle n’_i n’_j|\hat{H}_{DD}|n_i n_j\rangle=A_m \langle n’_i| \hat{d}_0^{(i)} |n_i\rangle \langle n’_j| \hat{d}_0^{(j)} |n_j\rangle \\
\end{aligned}
\end{equation}
where $A_m$ is the prefactor $\frac{1-3\cos^2\theta_{ij}}{r_{ij}^3}$, and $|n_i\rangle$ and $|n_j\rangle$ are the eigenstates of the $i$-th and $j$-th molecule defined above ($n_i,\,n_j=-1,0,1$). The ED Hamiltonian can be expressed in matrix form as
\begin{equation}
\begin{aligned} 
A_m
\begin{pmatrix}
 & |11\rangle & |10\rangle &|1-1\rangle &|01\rangle &|00\rangle &|0-1\rangle &|-11\rangle &|-10\rangle &|-1-1\rangle \\
\langle 11 |& d_1 d_1&0&0&0&0&0&0&0&0 \\
\langle 10| &0 &d_1 d_0&0&d_{10} d_{01}&0&d_{10} d_{0-1}&0&0&0\\
\langle 1-1| &0  &0&d_1 d_{-1}&0&0&0&0&0&0\\
\langle 01| &0 &d_{01} d_{10}&0&d_0 d_1&0&0&0&d_{0-1} d_{10}&0\\
\langle 00| &0 &0&0&0&d_0 d_0&0&0&0&0\\
\langle 0-1| &0 &d_{01} d_{-10}&0&0&0&d_0 d_{-1}&0&d_{0-1} d_{-10}&0\\
\langle -11| &0 &0&0&0&0&0&d_{-1} d_1&0&0\\
\langle -10| &0  &0&0&d_{-10}d_{01}&0&d_{-10} d_{0-1}&0&d_{-1} d_0&0\\
\langle-1-1 | &0 &0&0&0&0&0&0&0&d_{-1} d_{-1}\\
\end{pmatrix}
\end{aligned}
\end{equation}
where $d_x=\langle x|\hat{d}_0|x\rangle$ and $d_{xy}=\langle x|\hat{d}_0|y\rangle$ are single-molecule EDM matrix elements. It should be noted that state $1$ and $-1$ have the same energy, and we exclusively consider processes that conserve energy. Additionally, we assume that transitions between states $-1$ and $1$ are not possible.  The magnetic field is assumed to be tuned to the midpoint of the avoided crossing, where the transition EDMs coincide (see Fig.~6), i.e., $d_{01}=d_{-10}=d_t$. Lastly, we assume the electric field is zero and thus the diagonal dipole matrix elements vanish, $d_x=0$. This resulting Hamiltonian matrix takes the form
\begin{equation}
\begin{aligned} 
A_m
\begin{pmatrix}
 & |11\rangle & |10\rangle &|1-1\rangle &|01\rangle &|00\rangle &|0-1\rangle &|-11\rangle &|-10\rangle &|-1-1\rangle \\
\langle 11 |& 0&0&0&0&0&0&0&0&0 \\
\langle 10| &0 &0&0&d_t^2&0&d_t^2&0&0&0\\
\langle 1-1| &0  &0&0&0&0&0&0&0&0\\
\langle 01| &0 &d_t^2&0&0&0&0&0&d_t^2&0\\
\langle 00| &0 &0&0&0&0&0&0&0&0\\
\langle 0-1| &0 &d_t^2&0&0&0&0&0&d_t^2&0\\
\langle -11| &0 &0&0&0&0&0&0&0&0\\
\langle -10| &0  &0&0&d_t^2&0&d_t^2&0&0&0\\
\langle-1-1 | &0 &0&0&0&0&0&0&0&0\\
\end{pmatrix}
\end{aligned}
\end{equation}
To construct the Hamiltonian, we need an operator for molecule $i$ that can be represented as $\begin{pmatrix}
0&1&0\\
0&0&0\\
0&1&0
\end{pmatrix} $ and an operator for molecule $j$ that can be represented as $\begin{pmatrix}
0&0&0\\
1&0&1\\
0&0&0
\end{pmatrix}$. Additionally, we require an operator for molecule $i$ that can be represented as $\begin{pmatrix}
0&0&0\\
1&0&1\\
0&0&0
\end{pmatrix} $and an operator for molecule $j$ that can be represented as  $\begin{pmatrix}
0&1&0\\
0&0&0\\
0&1&0
\end{pmatrix}$.
We aim to express the aforementioned operators in the spin-1 basis using the the $S = 1$ spin projection operators $\hat{S}_\alpha$ ($\alpha = x,y,z$) and the quadrupole (nematic)  moments $\hat{Q}_{\alpha\beta}$. The relevant matrix representations of the $S=1$ operators and nematic moments are
\begin{equation}
\begin{aligned} 
S_x=\frac{1}{\sqrt2}
\begin{pmatrix}
0&1&0\\
1&0&1\\
0&1&0
\end{pmatrix}\\
Q_{yz}=(S_yS_z+S_zS_y)=\frac{1}{\sqrt2}
\begin{pmatrix}
0&-i&0\\
i&0&i\\
0&-i&0
\end{pmatrix}\\
\end{aligned}
\end{equation}
which can be combined to obtain 
\begin{equation}
\begin{aligned} 
S_x+iQ_{yz}=\frac{2}{\sqrt2}
\begin{pmatrix}
0&1&0\\
0&0&0\\
0&1&0
\end{pmatrix}\\
S_x-iQ_{yz}=\frac{2}{\sqrt2}
\begin{pmatrix}
0&0&0\\
1&0&1\\
0&0&0
\end{pmatrix}\\
\end{aligned}
\end{equation}

These operators can be used to obtain the Hamiltonian in terms of spin-1 operators 
\begin{equation}
\begin{aligned} 
\hat{H}_{DD}=A_m d_t^2\bigg(\frac{1}{\sqrt{2}}(\hat{S}_x^i+i\hat{Q}_{yz}^i)\otimes \frac{1}{\sqrt{2}}(\hat{S}_x^j-i\hat{Q}_{yz}^j)+\frac{1}{\sqrt{2}}(\hat{S}_x^i-i\hat{Q}_{yz}^i)\otimes \frac{1}{\sqrt{2}}(\hat{S}_x^j+i\hat{Q}_{yz}^j)\bigg)\\
=\frac{d^2_t}{2}\frac{1- 3\cos^2\theta_{ij}}{R_{ij}^3} \bigg( (\hat{S}_x^i+i\hat{Q}_{yz}^i)\otimes (\hat{S}_x^j-i\hat{Q}_{yz}^j)+(\hat{S}_x^i-i\hat{Q}_{yz}^i)\otimes(\hat{S}_x^j+i\hat{Q}_{yz}^j)\bigg)
\end{aligned}
\end{equation}

Notably, this Hamiltonian is identical to the one derived in section V, thereby affirming our findings. The Hamiltonian can be further simplified to obtain
\begin{equation}
\begin{aligned} 
\hat{H}_{DD}=
\frac{1}{2} d_t^2 \frac{1- 3\cos^2\theta_{ij}}{R_{ij}^3}\bigg((\hat{S}_x^i \otimes \hat{S}_x^j+i\hat{Q}_{yz}^i \otimes \hat{S}_x^j-i\hat{S}_x^i\otimes \hat{Q}_{yz}^j+\hat{Q}_{yz}^i\otimes \hat{Q}_{yz}^j)\\
+(\hat{S}_x^i \otimes \hat{S}_x^j-i\hat{Q}_{yz}^i \otimes \hat{S}_x^j+i\hat{S}_x^i\otimes \hat{Q}_{yz}^j+\hat{Q}_{yz}^i\otimes \hat{Q}_{yz}^j)\bigg)\\
= d_t^2 \frac{1- 3\cos^2\theta_{ij}}{R_{ij}^3} \bigg(\hat{S}_x^i \otimes \hat{S}_x^j+\hat{Q}_{yz}^i\otimes \hat{Q}_{yz}^j\bigg).
\end{aligned}
\end{equation}

\bibliography{ColdMol2,cold_mol}

\begin{thebibliography}{74}%
\makeatletter
\providecommand \@ifxundefined [1]{%
 \@ifx{#1\undefined}
}%
\providecommand \@ifnum [1]{%
 \ifnum #1\expandafter \@firstoftwo
 \else \expandafter \@secondoftwo
 \fi
}%
\providecommand \@ifx [1]{%
 \ifx #1\expandafter \@firstoftwo
 \else \expandafter \@secondoftwo
 \fi
}%
\providecommand \natexlab [1]{#1}%
\providecommand \enquote  [1]{``#1''}%
\providecommand \bibnamefont  [1]{#1}%
\providecommand \bibfnamefont [1]{#1}%
\providecommand \citenamefont [1]{#1}%
\providecommand \href@noop [0]{\@secondoftwo}%
\providecommand \href [0]{\begingroup \@sanitize@url \@href}%
\providecommand \@href[1]{\@@startlink{#1}\@@href}%
\providecommand \@@href[1]{\endgroup#1\@@endlink}%
\providecommand \@sanitize@url [0]{\catcode `\\12\catcode `\$12\catcode
  `\&12\catcode `\#12\catcode `\^12\catcode `\_12\catcode `\%12\relax}%
\providecommand \@@startlink[1]{}%
\providecommand \@@endlink[0]{}%
\providecommand \url  [0]{\begingroup\@sanitize@url \@url }%
\providecommand \@url [1]{\endgroup\@href {#1}{\urlprefix }}%
\providecommand \urlprefix  [0]{URL }%
\providecommand \Eprint [0]{\href }%
\providecommand \doibase [0]{https://doi.org/}%
\providecommand \selectlanguage [0]{\@gobble}%
\providecommand \bibinfo  [0]{\@secondoftwo}%
\providecommand \bibfield  [0]{\@secondoftwo}%
\providecommand \translation [1]{[#1]}%
\providecommand \BibitemOpen [0]{}%
\providecommand \bibitemStop [0]{}%
\providecommand \bibitemNoStop [0]{.\EOS\space}%
\providecommand \EOS [0]{\spacefactor3000\relax}%
\providecommand \BibitemShut  [1]{\csname bibitem#1\endcsname}%
\let\auto@bib@innerbib\@empty
\bibitem [{\citenamefont {DeMille}(2002)}]{DeMille:02}%
  \BibitemOpen
  \bibfield  {author} {\bibinfo {author} {\bibfnamefont {D.}~\bibnamefont
  {DeMille}},\ }\bibfield  {title} {\bibinfo {title} {Quantum computation with
  trapped polar molecules},\ }\href@noop {} {\bibfield  {journal} {\bibinfo
  {journal} {Phys. Lev. Lett.}\ }\textbf {\bibinfo {volume} {88}} (\bibinfo
  {year} {2002})}\BibitemShut {NoStop}%
\bibitem [{\citenamefont {Yelin}\ \emph {et~al.}(2006)\citenamefont {Yelin},
  \citenamefont {Kirby},\ and\ \citenamefont {C\^ot\'e}}]{Yelin:06}%
  \BibitemOpen
  \bibfield  {author} {\bibinfo {author} {\bibfnamefont {S.~F.}\ \bibnamefont
  {Yelin}}, \bibinfo {author} {\bibfnamefont {K.}~\bibnamefont {Kirby}},\ and\
  \bibinfo {author} {\bibfnamefont {R.}~\bibnamefont {C\^ot\'e}},\ }\bibfield
  {title} {\bibinfo {title} {Schemes for robust quantum computation with polar
  molecules},\ }\href {https://doi.org/10.1103/PhysRevA.74.050301} {\bibfield
  {journal} {\bibinfo  {journal} {Phys. Rev. A}\ }\textbf {\bibinfo {volume}
  {74}},\ \bibinfo {pages} {050301} (\bibinfo {year} {2006})}\BibitemShut
  {NoStop}%
\bibitem [{\citenamefont {Yan}\ \emph {et~al.}(2013)\citenamefont {Yan},
  \citenamefont {Moses}, \citenamefont {Gadway}, \citenamefont {Covey},
  \citenamefont {Hazzard}, \citenamefont {Rey}, \citenamefont {Jin},\ and\
  \citenamefont {Ye}}]{Yan:13}%
  \BibitemOpen
  \bibfield  {author} {\bibinfo {author} {\bibfnamefont {B.}~\bibnamefont
  {Yan}}, \bibinfo {author} {\bibfnamefont {S.~A.}\ \bibnamefont {Moses}},
  \bibinfo {author} {\bibfnamefont {B.}~\bibnamefont {Gadway}}, \bibinfo
  {author} {\bibfnamefont {J.~P.}\ \bibnamefont {Covey}}, \bibinfo {author}
  {\bibfnamefont {K.~R.~A.}\ \bibnamefont {Hazzard}}, \bibinfo {author}
  {\bibfnamefont {A.~M.}\ \bibnamefont {Rey}}, \bibinfo {author} {\bibfnamefont
  {D.~S.}\ \bibnamefont {Jin}},\ and\ \bibinfo {author} {\bibfnamefont
  {J.}~\bibnamefont {Ye}},\ }\bibfield  {title} {\bibinfo {title} {Observation
  of dipolar spin-exchange interactions with lattice-confined polar
  molecules},\ }\href {https://doi.org/10.1038/nature12483} {\bibfield
  {journal} {\bibinfo  {journal} {Nature}\ }\textbf {\bibinfo {volume} {501}},\
  \bibinfo {pages} {521} (\bibinfo {year} {2013})}\BibitemShut {NoStop}%
\bibitem [{\citenamefont {Bohn}\ \emph {et~al.}(2017)\citenamefont {Bohn},
  \citenamefont {Rey},\ and\ \citenamefont {Ye}}]{Bohn:17}%
  \BibitemOpen
  \bibfield  {author} {\bibinfo {author} {\bibfnamefont {J.~L.}\ \bibnamefont
  {Bohn}}, \bibinfo {author} {\bibfnamefont {A.~M.}\ \bibnamefont {Rey}},\ and\
  \bibinfo {author} {\bibfnamefont {J.}~\bibnamefont {Ye}},\ }\bibfield
  {title} {\bibinfo {title} {Cold molecules: Progress in quantum engineering of
  chemistry and quantum matter},\ }\href@noop {} {\bibfield  {journal}
  {\bibinfo  {journal} {Science}\ }\textbf {\bibinfo {volume} {357}},\ \bibinfo
  {pages} {1002} (\bibinfo {year} {2017})}\BibitemShut {NoStop}%
\bibitem [{\citenamefont {Albert}\ \emph {et~al.}(2020)\citenamefont {Albert},
  \citenamefont {Covey},\ and\ \citenamefont {Preskill}}]{Albert:20}%
  \BibitemOpen
  \bibfield  {author} {\bibinfo {author} {\bibfnamefont {V.~V.}\ \bibnamefont
  {Albert}}, \bibinfo {author} {\bibfnamefont {J.~P.}\ \bibnamefont {Covey}},\
  and\ \bibinfo {author} {\bibfnamefont {J.}~\bibnamefont {Preskill}},\
  }\bibfield  {title} {\bibinfo {title} {Robust encoding of a qubit in a
  molecule},\ }\href {https://doi.org/10.1103/PhysRevX.10.031050} {\bibfield
  {journal} {\bibinfo  {journal} {Phys. Rev. X}\ }\textbf {\bibinfo {volume}
  {10}},\ \bibinfo {pages} {031050} (\bibinfo {year} {2020})}\BibitemShut
  {NoStop}%
\bibitem [{\citenamefont {Park}\ \emph {et~al.}(2017)\citenamefont {Park},
  \citenamefont {Yan}, \citenamefont {Loh}, \citenamefont {Will},\ and\
  \citenamefont {Zwierlein}}]{Park:17}%
  \BibitemOpen
  \bibfield  {author} {\bibinfo {author} {\bibfnamefont {J.~W.}\ \bibnamefont
  {Park}}, \bibinfo {author} {\bibfnamefont {Z.~Z.}\ \bibnamefont {Yan}},
  \bibinfo {author} {\bibfnamefont {H.}~\bibnamefont {Loh}}, \bibinfo {author}
  {\bibfnamefont {S.~A.}\ \bibnamefont {Will}},\ and\ \bibinfo {author}
  {\bibfnamefont {M.~W.}\ \bibnamefont {Zwierlein}},\ }\bibfield  {title}
  {\bibinfo {title} {Second-scale nuclear spin coherence time of ultracold
  {$^{23}$Na$^{40}$K} molecules},\ }\href
  {https://doi.org/10.1126/science.aal5066} {\bibfield  {journal} {\bibinfo
  {journal} {Science}\ }\textbf {\bibinfo {volume} {357}},\ \bibinfo {pages}
  {372} (\bibinfo {year} {2017})}\BibitemShut {NoStop}%
\bibitem [{\citenamefont {Li}\ \emph {et~al.}(2023)\citenamefont {Li},
  \citenamefont {Matsuda}, \citenamefont {Miller}, \citenamefont {Carroll},
  \citenamefont {Tobias}, \citenamefont {Higgins},\ and\ \citenamefont
  {Ye}}]{Li:23}%
  \BibitemOpen
  \bibfield  {author} {\bibinfo {author} {\bibfnamefont {J.-R.}\ \bibnamefont
  {Li}}, \bibinfo {author} {\bibfnamefont {K.}~\bibnamefont {Matsuda}},
  \bibinfo {author} {\bibfnamefont {C.}~\bibnamefont {Miller}}, \bibinfo
  {author} {\bibfnamefont {A.~N.}\ \bibnamefont {Carroll}}, \bibinfo {author}
  {\bibfnamefont {W.~G.}\ \bibnamefont {Tobias}}, \bibinfo {author}
  {\bibfnamefont {J.~S.}\ \bibnamefont {Higgins}},\ and\ \bibinfo {author}
  {\bibfnamefont {J.}~\bibnamefont {Ye}},\ }\bibfield  {title} {\bibinfo
  {title} {Tunable itinerant spin dynamics with polar molecules},\ }\href
  {https://doi.org/10.1038/s41586-022-05479-2} {\bibfield  {journal} {\bibinfo
  {journal} {Nature}\ }\textbf {\bibinfo {volume} {614}},\ \bibinfo {pages}
  {70} (\bibinfo {year} {2023})}\BibitemShut {NoStop}%
\bibitem [{\citenamefont {Tobias}\ \emph {et~al.}(2022)\citenamefont {Tobias},
  \citenamefont {Matsuda}, \citenamefont {Li}, \citenamefont {Miller},
  \citenamefont {Carroll}, \citenamefont {Bilitewski}, \citenamefont {Rey},\
  and\ \citenamefont {Ye}}]{Tobias:22}%
  \BibitemOpen
  \bibfield  {author} {\bibinfo {author} {\bibfnamefont {W.~G.}\ \bibnamefont
  {Tobias}}, \bibinfo {author} {\bibfnamefont {K.}~\bibnamefont {Matsuda}},
  \bibinfo {author} {\bibfnamefont {J.-R.}\ \bibnamefont {Li}}, \bibinfo
  {author} {\bibfnamefont {C.}~\bibnamefont {Miller}}, \bibinfo {author}
  {\bibfnamefont {A.~N.}\ \bibnamefont {Carroll}}, \bibinfo {author}
  {\bibfnamefont {T.}~\bibnamefont {Bilitewski}}, \bibinfo {author}
  {\bibfnamefont {A.~M.}\ \bibnamefont {Rey}},\ and\ \bibinfo {author}
  {\bibfnamefont {J.}~\bibnamefont {Ye}},\ }\bibfield  {title} {\bibinfo
  {title} {Reactions between layer-resolved molecules mediated by dipolar spin
  exchange},\ }\href {https://doi.org/10.1126/science.abn8525} {\bibfield
  {journal} {\bibinfo  {journal} {Science}\ }\textbf {\bibinfo {volume}
  {375}},\ \bibinfo {pages} {1299} (\bibinfo {year} {2022})}\BibitemShut
  {NoStop}%
\bibitem [{\citenamefont {Christakis}\ \emph {et~al.}(2023)\citenamefont
  {Christakis}, \citenamefont {Rosenberg}, \citenamefont {Raj}, \citenamefont
  {Chi}, \citenamefont {Morningstar}, \citenamefont {Huse}, \citenamefont
  {Yan},\ and\ \citenamefont {Bakr}}]{Christakis:23}%
  \BibitemOpen
  \bibfield  {author} {\bibinfo {author} {\bibfnamefont {L.}~\bibnamefont
  {Christakis}}, \bibinfo {author} {\bibfnamefont {J.~S.}\ \bibnamefont
  {Rosenberg}}, \bibinfo {author} {\bibfnamefont {R.}~\bibnamefont {Raj}},
  \bibinfo {author} {\bibfnamefont {S.}~\bibnamefont {Chi}}, \bibinfo {author}
  {\bibfnamefont {A.}~\bibnamefont {Morningstar}}, \bibinfo {author}
  {\bibfnamefont {D.~A.}\ \bibnamefont {Huse}}, \bibinfo {author}
  {\bibfnamefont {Z.~Z.}\ \bibnamefont {Yan}},\ and\ \bibinfo {author}
  {\bibfnamefont {W.~S.}\ \bibnamefont {Bakr}},\ }\bibfield  {title} {\bibinfo
  {title} {Probing site-resolved correlations in a spin system of ultracold
  molecules},\ }\href {https://doi.org/10.1038/s41586-022-05558-4} {\bibfield
  {journal} {\bibinfo  {journal} {Nature}\ }\textbf {\bibinfo {volume} {614}},\
  \bibinfo {pages} {64} (\bibinfo {year} {2023})}\BibitemShut {NoStop}%
\bibitem [{\citenamefont {Zhang}\ \emph {et~al.}(2020)\citenamefont {Zhang},
  \citenamefont {Yu}, \citenamefont {Cairncross}, \citenamefont {Wang},
  \citenamefont {Picard}, \citenamefont {Hood}, \citenamefont {Lin},
  \citenamefont {Hutson},\ and\ \citenamefont {Ni}}]{Zhang:20}%
  \BibitemOpen
  \bibfield  {author} {\bibinfo {author} {\bibfnamefont {J.~T.}\ \bibnamefont
  {Zhang}}, \bibinfo {author} {\bibfnamefont {Y.}~\bibnamefont {Yu}}, \bibinfo
  {author} {\bibfnamefont {W.~B.}\ \bibnamefont {Cairncross}}, \bibinfo
  {author} {\bibfnamefont {K.}~\bibnamefont {Wang}}, \bibinfo {author}
  {\bibfnamefont {L.~R.~B.}\ \bibnamefont {Picard}}, \bibinfo {author}
  {\bibfnamefont {J.~D.}\ \bibnamefont {Hood}}, \bibinfo {author}
  {\bibfnamefont {Y.-W.}\ \bibnamefont {Lin}}, \bibinfo {author} {\bibfnamefont
  {J.~M.}\ \bibnamefont {Hutson}},\ and\ \bibinfo {author} {\bibfnamefont
  {K.-K.}\ \bibnamefont {Ni}},\ }\bibfield  {title} {\bibinfo {title} {Forming
  a single molecule by magnetoassociation in an optical tweezer},\ }\href
  {https://doi.org/10.1103/PhysRevLett.124.253401} {\bibfield  {journal}
  {\bibinfo  {journal} {Phys. Rev. Lett.}\ }\textbf {\bibinfo {volume} {124}},\
  \bibinfo {pages} {253401} (\bibinfo {year} {2020})}\BibitemShut {NoStop}%
\bibitem [{\citenamefont {He}\ \emph {et~al.}(2020)\citenamefont {He},
  \citenamefont {Wang}, \citenamefont {Zhuang}, \citenamefont {Xu},
  \citenamefont {Gao}, \citenamefont {Guo}, \citenamefont {Sheng},
  \citenamefont {Liu}, \citenamefont {Wang}, \citenamefont {Li}, \citenamefont
  {Shlyapnikov},\ and\ \citenamefont {Zhan}}]{He:20}%
  \BibitemOpen
  \bibfield  {author} {\bibinfo {author} {\bibfnamefont {X.}~\bibnamefont
  {He}}, \bibinfo {author} {\bibfnamefont {K.}~\bibnamefont {Wang}}, \bibinfo
  {author} {\bibfnamefont {J.}~\bibnamefont {Zhuang}}, \bibinfo {author}
  {\bibfnamefont {P.}~\bibnamefont {Xu}}, \bibinfo {author} {\bibfnamefont
  {X.}~\bibnamefont {Gao}}, \bibinfo {author} {\bibfnamefont {R.}~\bibnamefont
  {Guo}}, \bibinfo {author} {\bibfnamefont {C.}~\bibnamefont {Sheng}}, \bibinfo
  {author} {\bibfnamefont {M.}~\bibnamefont {Liu}}, \bibinfo {author}
  {\bibfnamefont {J.}~\bibnamefont {Wang}}, \bibinfo {author} {\bibfnamefont
  {J.}~\bibnamefont {Li}}, \bibinfo {author} {\bibfnamefont {G.~V.}\
  \bibnamefont {Shlyapnikov}},\ and\ \bibinfo {author} {\bibfnamefont
  {M.}~\bibnamefont {Zhan}},\ }\bibfield  {title} {\bibinfo {title} {Coherently
  forming a single molecule in an optical trap},\ }\href
  {https://doi.org/10.1126/science.aba7468} {\bibfield  {journal} {\bibinfo
  {journal} {Science}\ }\textbf {\bibinfo {volume} {370}},\ \bibinfo {pages}
  {331} (\bibinfo {year} {2020})}\BibitemShut {NoStop}%
\bibitem [{\citenamefont {Cairncross}\ \emph {et~al.}(2021)\citenamefont
  {Cairncross}, \citenamefont {Zhang}, \citenamefont {Picard}, \citenamefont
  {Yu}, \citenamefont {Wang},\ and\ \citenamefont {Ni}}]{Cairncross:21}%
  \BibitemOpen
  \bibfield  {author} {\bibinfo {author} {\bibfnamefont {W.~B.}\ \bibnamefont
  {Cairncross}}, \bibinfo {author} {\bibfnamefont {J.~T.}\ \bibnamefont
  {Zhang}}, \bibinfo {author} {\bibfnamefont {L.~R.~B.}\ \bibnamefont
  {Picard}}, \bibinfo {author} {\bibfnamefont {Y.}~\bibnamefont {Yu}}, \bibinfo
  {author} {\bibfnamefont {K.}~\bibnamefont {Wang}},\ and\ \bibinfo {author}
  {\bibfnamefont {K.-K.}\ \bibnamefont {Ni}},\ }\bibfield  {title} {\bibinfo
  {title} {Assembly of a rovibrational ground state molecule in an optical
  tweezer},\ }\href {https://doi.org/10.1103/PhysRevLett.126.123402} {\bibfield
   {journal} {\bibinfo  {journal} {Phys. Rev. Lett.}\ }\textbf {\bibinfo
  {volume} {126}},\ \bibinfo {pages} {123402} (\bibinfo {year}
  {2021})}\BibitemShut {NoStop}%
\bibitem [{\citenamefont {Zhang}\ \emph {et~al.}(2022)\citenamefont {Zhang},
  \citenamefont {Picard}, \citenamefont {Cairncross}, \citenamefont {Wang},
  \citenamefont {Yu}, \citenamefont {Fang},\ and\ \citenamefont
  {Ni}}]{zhang2022optical}%
  \BibitemOpen
  \bibfield  {author} {\bibinfo {author} {\bibfnamefont {J.~T.}\ \bibnamefont
  {Zhang}}, \bibinfo {author} {\bibfnamefont {L.~R.~B.}\ \bibnamefont
  {Picard}}, \bibinfo {author} {\bibfnamefont {W.~B.}\ \bibnamefont
  {Cairncross}}, \bibinfo {author} {\bibfnamefont {K.}~\bibnamefont {Wang}},
  \bibinfo {author} {\bibfnamefont {Y.}~\bibnamefont {Yu}}, \bibinfo {author}
  {\bibfnamefont {F.}~\bibnamefont {Fang}},\ and\ \bibinfo {author}
  {\bibfnamefont {K.-K.}\ \bibnamefont {Ni}},\ }\bibfield  {title} {\bibinfo
  {title} {An optical tweezer array of ground-state polar molecules},\ }\href
  {https://doi.org/10.1088/2058-9565/ac676c} {\bibfield  {journal} {\bibinfo
  {journal} {Quantum Sci. Technol.}\ }\textbf {\bibinfo {volume} {7}},\
  \bibinfo {pages} {035006} (\bibinfo {year} {2022})}\BibitemShut {NoStop}%
\bibitem [{\citenamefont {Burchesky}\ \emph {et~al.}(2021)\citenamefont
  {Burchesky}, \citenamefont {Anderegg}, \citenamefont {Bao}, \citenamefont
  {Yu}, \citenamefont {Chae}, \citenamefont {Ketterle}, \citenamefont {Ni},\
  and\ \citenamefont {Doyle}}]{Burchesky:21}%
  \BibitemOpen
  \bibfield  {author} {\bibinfo {author} {\bibfnamefont {S.}~\bibnamefont
  {Burchesky}}, \bibinfo {author} {\bibfnamefont {L.}~\bibnamefont {Anderegg}},
  \bibinfo {author} {\bibfnamefont {Y.}~\bibnamefont {Bao}}, \bibinfo {author}
  {\bibfnamefont {S.~S.}\ \bibnamefont {Yu}}, \bibinfo {author} {\bibfnamefont
  {E.}~\bibnamefont {Chae}}, \bibinfo {author} {\bibfnamefont {W.}~\bibnamefont
  {Ketterle}}, \bibinfo {author} {\bibfnamefont {K.-K.}\ \bibnamefont {Ni}},\
  and\ \bibinfo {author} {\bibfnamefont {J.~M.}\ \bibnamefont {Doyle}},\
  }\bibfield  {title} {\bibinfo {title} {Rotational coherence times of polar
  molecules in optical tweezers},\ }\href
  {https://doi.org/10.1103/PhysRevLett.127.123202} {\bibfield  {journal}
  {\bibinfo  {journal} {Phys. Rev. Lett.}\ }\textbf {\bibinfo {volume} {127}},\
  \bibinfo {pages} {123202} (\bibinfo {year} {2021})}\BibitemShut {NoStop}%
\bibitem [{\citenamefont {Holland}\ \emph {et~al.}(2022)\citenamefont
  {Holland}, \citenamefont {Lu},\ and\ \citenamefont
  {Cheuk}}]{holland2022demand}%
  \BibitemOpen
  \bibfield  {author} {\bibinfo {author} {\bibfnamefont {C.~M.}\ \bibnamefont
  {Holland}}, \bibinfo {author} {\bibfnamefont {Y.}~\bibnamefont {Lu}},\ and\
  \bibinfo {author} {\bibfnamefont {L.~W.}\ \bibnamefont {Cheuk}},\ }\bibfield
  {title} {\bibinfo {title} {On-demand entanglement of molecules in a
  reconfigurable optical tweezer array},\ }\href@noop {} {\bibfield  {journal}
  {\bibinfo  {journal} {arXiv preprint arXiv:2210.06309}\ } (\bibinfo {year}
  {2022})}\BibitemShut {NoStop}%
\bibitem [{\citenamefont {Bao}\ \emph {et~al.}(2022)\citenamefont {Bao},
  \citenamefont {Yu}, \citenamefont {Anderegg}, \citenamefont {Chae},
  \citenamefont {Ketterle}, \citenamefont {Ni},\ and\ \citenamefont
  {Doyle}}]{bao2022dipolar}%
  \BibitemOpen
  \bibfield  {author} {\bibinfo {author} {\bibfnamefont {Y.}~\bibnamefont
  {Bao}}, \bibinfo {author} {\bibfnamefont {S.~S.}\ \bibnamefont {Yu}},
  \bibinfo {author} {\bibfnamefont {L.}~\bibnamefont {Anderegg}}, \bibinfo
  {author} {\bibfnamefont {E.}~\bibnamefont {Chae}}, \bibinfo {author}
  {\bibfnamefont {W.}~\bibnamefont {Ketterle}}, \bibinfo {author}
  {\bibfnamefont {K.-K.}\ \bibnamefont {Ni}},\ and\ \bibinfo {author}
  {\bibfnamefont {J.~M.}\ \bibnamefont {Doyle}},\ }\bibfield  {title} {\bibinfo
  {title} {Dipolar spin-exchange and entanglement between molecules in an
  optical tweezer array},\ }\href@noop {} {\bibfield  {journal} {\bibinfo
  {journal} {arXiv preprint arXiv:2211.09780}\ } (\bibinfo {year}
  {2022})}\BibitemShut {NoStop}%
\bibitem [{\citenamefont {Krems}(2008)}]{Krems:08}%
  \BibitemOpen
  \bibfield  {author} {\bibinfo {author} {\bibfnamefont {R.~V.}\ \bibnamefont
  {Krems}},\ }\bibfield  {title} {\bibinfo {title} {Cold controlled
  chemistry},\ }\href@noop {} {\bibfield  {journal} {\bibinfo  {journal} {Phys.
  Chem. Chem. Phys.}\ }\textbf {\bibinfo {volume} {10}},\ \bibinfo {pages}
  {4079} (\bibinfo {year} {2008})}\BibitemShut {NoStop}%
\bibitem [{\citenamefont {Balakrishnan}(2016)}]{Balakrishnan:16}%
  \BibitemOpen
  \bibfield  {author} {\bibinfo {author} {\bibfnamefont {N.}~\bibnamefont
  {Balakrishnan}},\ }\bibfield  {title} {\bibinfo {title} {Perspective:
  Ultracold molecules and the dawn of cold controlled chemistry},\ }\href
  {https://doi.org/10.1063/1.4964096} {\bibfield  {journal} {\bibinfo
  {journal} {J. Chem. Phys.}\ }\textbf {\bibinfo {volume} {145}},\ \bibinfo
  {pages} {150901} (\bibinfo {year} {2016})}\BibitemShut {NoStop}%
\bibitem [{\citenamefont {DeMille}\ \emph {et~al.}(2017)\citenamefont
  {DeMille}, \citenamefont {Doyle},\ and\ \citenamefont
  {Sushkov}}]{DeMille:17}%
  \BibitemOpen
  \bibfield  {author} {\bibinfo {author} {\bibfnamefont {D.}~\bibnamefont
  {DeMille}}, \bibinfo {author} {\bibfnamefont {J.~M.}\ \bibnamefont {Doyle}},\
  and\ \bibinfo {author} {\bibfnamefont {A.~O.}\ \bibnamefont {Sushkov}},\
  }\bibfield  {title} {\bibinfo {title} {Probing the frontiers of particle
  physics with tabletop-scale experiments},\ }\href
  {https://doi.org/10.1126/science.aal3003} {\bibfield  {journal} {\bibinfo
  {journal} {Science}\ }\textbf {\bibinfo {volume} {357}},\ \bibinfo {pages}
  {990} (\bibinfo {year} {2017})}\BibitemShut {NoStop}%
\bibitem [{\citenamefont {Ni}\ \emph {et~al.}(2018)\citenamefont {Ni},
  \citenamefont {Rosenband},\ and\ \citenamefont {Grimes}}]{Ni:18}%
  \BibitemOpen
  \bibfield  {author} {\bibinfo {author} {\bibfnamefont {K.-K.}\ \bibnamefont
  {Ni}}, \bibinfo {author} {\bibfnamefont {T.}~\bibnamefont {Rosenband}},\ and\
  \bibinfo {author} {\bibfnamefont {D.~D.}\ \bibnamefont {Grimes}},\ }\bibfield
   {title} {\bibinfo {title} {Dipolar exchange quantum logic gate with polar
  molecules},\ }\href {https://doi.org/10.1039/C8SC02355G} {\bibfield
  {journal} {\bibinfo  {journal} {Chem. Sci.}\ }\textbf {\bibinfo {volume}
  {9}},\ \bibinfo {pages} {6830} (\bibinfo {year} {2018})}\BibitemShut
  {NoStop}%
\bibitem [{\citenamefont {Pezz\`e}\ \emph {et~al.}(2018)\citenamefont
  {Pezz\`e}, \citenamefont {Smerzi}, \citenamefont {Oberthaler}, \citenamefont
  {Schmied},\ and\ \citenamefont {Treutlein}}]{Pezze:18}%
  \BibitemOpen
  \bibfield  {author} {\bibinfo {author} {\bibfnamefont {L.}~\bibnamefont
  {Pezz\`e}}, \bibinfo {author} {\bibfnamefont {A.}~\bibnamefont {Smerzi}},
  \bibinfo {author} {\bibfnamefont {M.~K.}\ \bibnamefont {Oberthaler}},
  \bibinfo {author} {\bibfnamefont {R.}~\bibnamefont {Schmied}},\ and\ \bibinfo
  {author} {\bibfnamefont {P.}~\bibnamefont {Treutlein}},\ }\bibfield  {title}
  {\bibinfo {title} {Quantum metrology with nonclassical states of atomic
  ensembles},\ }\href {https://doi.org/10.1103/RevModPhys.90.035005} {\bibfield
   {journal} {\bibinfo  {journal} {Rev. Mod. Phys.}\ }\textbf {\bibinfo
  {volume} {90}},\ \bibinfo {pages} {035005} (\bibinfo {year}
  {2018})}\BibitemShut {NoStop}%
\bibitem [{\citenamefont {Bilitewski}\ \emph {et~al.}(2021)\citenamefont
  {Bilitewski}, \citenamefont {De~Marco}, \citenamefont {Li}, \citenamefont
  {Matsuda}, \citenamefont {Tobias}, \citenamefont {Valtolina}, \citenamefont
  {Ye},\ and\ \citenamefont {Rey}}]{Bilitewski:21}%
  \BibitemOpen
  \bibfield  {author} {\bibinfo {author} {\bibfnamefont {T.}~\bibnamefont
  {Bilitewski}}, \bibinfo {author} {\bibfnamefont {L.}~\bibnamefont
  {De~Marco}}, \bibinfo {author} {\bibfnamefont {J.-R.}\ \bibnamefont {Li}},
  \bibinfo {author} {\bibfnamefont {K.}~\bibnamefont {Matsuda}}, \bibinfo
  {author} {\bibfnamefont {W.~G.}\ \bibnamefont {Tobias}}, \bibinfo {author}
  {\bibfnamefont {G.}~\bibnamefont {Valtolina}}, \bibinfo {author}
  {\bibfnamefont {J.}~\bibnamefont {Ye}},\ and\ \bibinfo {author}
  {\bibfnamefont {A.~M.}\ \bibnamefont {Rey}},\ }\bibfield  {title} {\bibinfo
  {title} {Dynamical generation of spin squeezing in ultracold dipolar
  molecules},\ }\href {https://doi.org/10.1103/PhysRevLett.126.113401}
  {\bibfield  {journal} {\bibinfo  {journal} {Phys. Rev. Lett.}\ }\textbf
  {\bibinfo {volume} {126}},\ \bibinfo {pages} {113401} (\bibinfo {year}
  {2021})}\BibitemShut {NoStop}%
\bibitem [{\citenamefont {Tscherbul}\ \emph {et~al.}(2023)\citenamefont
  {Tscherbul}, \citenamefont {Ye},\ and\ \citenamefont {Rey}}]{Tscherbul:23}%
  \BibitemOpen
  \bibfield  {author} {\bibinfo {author} {\bibfnamefont {T.~V.}\ \bibnamefont
  {Tscherbul}}, \bibinfo {author} {\bibfnamefont {J.}~\bibnamefont {Ye}},\ and\
  \bibinfo {author} {\bibfnamefont {A.~M.}\ \bibnamefont {Rey}},\ }\bibfield
  {title} {\bibinfo {title} {Robust nuclear spin entanglement via dipolar
  interactions in polar molecules},\ }\href
  {https://doi.org/10.1103/PhysRevLett.130.143002} {\bibfield  {journal}
  {\bibinfo  {journal} {Phys. Rev. Lett.}\ }\textbf {\bibinfo {volume} {130}},\
  \bibinfo {pages} {143002} (\bibinfo {year} {2023})}\BibitemShut {NoStop}%
\bibitem [{\citenamefont {Bilitewski}\ and\ \citenamefont
  {Rey}(2023)}]{Bilitewski:23}%
  \BibitemOpen
  \bibfield  {author} {\bibinfo {author} {\bibfnamefont {T.}~\bibnamefont
  {Bilitewski}}\ and\ \bibinfo {author} {\bibfnamefont {A.~M.}\ \bibnamefont
  {Rey}},\ }\bibfield  {title} {\bibinfo {title} {Manipulating growth and
  propagation of correlations in dipolar multilayers: From pair production to
  bosonic {Kitaev} models},\ }\href
  {https://doi.org/10.1103/PhysRevLett.131.053001} {\bibfield  {journal}
  {\bibinfo  {journal} {Phys. Rev. Lett.}\ }\textbf {\bibinfo {volume} {131}},\
  \bibinfo {pages} {053001} (\bibinfo {year} {2023})}\BibitemShut {NoStop}%
\bibitem [{\citenamefont {Micheli}\ \emph {et~al.}(2006)\citenamefont
  {Micheli}, \citenamefont {Brennen},\ and\ \citenamefont
  {Zoller}}]{micheli2006toolbox}%
  \BibitemOpen
  \bibfield  {author} {\bibinfo {author} {\bibfnamefont {A.}~\bibnamefont
  {Micheli}}, \bibinfo {author} {\bibfnamefont {G.~K.}\ \bibnamefont
  {Brennen}},\ and\ \bibinfo {author} {\bibfnamefont {P.}~\bibnamefont
  {Zoller}},\ }\bibfield  {title} {\bibinfo {title} {A toolbox for lattice-spin
  models with polar molecules},\ }\href@noop {} {\bibfield  {journal} {\bibinfo
   {journal} {Nat. Phys.}\ }\textbf {\bibinfo {volume} {2}},\ \bibinfo {pages}
  {341} (\bibinfo {year} {2006})}\BibitemShut {NoStop}%
\bibitem [{\citenamefont {Micheli}\ \emph {et~al.}(2007)\citenamefont
  {Micheli}, \citenamefont {Pupillo}, \citenamefont {B\"uchler},\ and\
  \citenamefont {Zoller}}]{Micheli:07}%
  \BibitemOpen
  \bibfield  {author} {\bibinfo {author} {\bibfnamefont {A.}~\bibnamefont
  {Micheli}}, \bibinfo {author} {\bibfnamefont {G.}~\bibnamefont {Pupillo}},
  \bibinfo {author} {\bibfnamefont {H.~P.}\ \bibnamefont {B\"uchler}},\ and\
  \bibinfo {author} {\bibfnamefont {P.}~\bibnamefont {Zoller}},\ }\bibfield
  {title} {\bibinfo {title} {Cold polar molecules in two-dimensional traps:
  Tailoring interactions with external fields for novel quantum phases},\
  }\href {https://doi.org/10.1103/PhysRevA.76.043604} {\bibfield  {journal}
  {\bibinfo  {journal} {Phys. Rev. A}\ }\textbf {\bibinfo {volume} {76}},\
  \bibinfo {pages} {043604} (\bibinfo {year} {2007})}\BibitemShut {NoStop}%
\bibitem [{\citenamefont {Carr}\ \emph {et~al.}(2009)\citenamefont {Carr},
  \citenamefont {DeMille}, \citenamefont {Krems},\ and\ \citenamefont
  {Ye}}]{Carr:09}%
  \BibitemOpen
  \bibfield  {author} {\bibinfo {author} {\bibfnamefont {L.~D.}\ \bibnamefont
  {Carr}}, \bibinfo {author} {\bibfnamefont {D.}~\bibnamefont {DeMille}},
  \bibinfo {author} {\bibfnamefont {R.~V.}\ \bibnamefont {Krems}},\ and\
  \bibinfo {author} {\bibfnamefont {J.}~\bibnamefont {Ye}},\ }\bibfield
  {title} {\bibinfo {title} {Cold and ultracold molecules: science, technology
  and applications},\ }\href {https://doi.org/10.1088/1367-2630/11/5/055049}
  {\bibfield  {journal} {\bibinfo  {journal} {New J. Phys.}\ }\textbf {\bibinfo
  {volume} {11}},\ \bibinfo {pages} {055049} (\bibinfo {year}
  {2009})}\BibitemShut {NoStop}%
\bibitem [{\citenamefont {Gorshkov}\ \emph {et~al.}(2011)\citenamefont
  {Gorshkov}, \citenamefont {Manmana}, \citenamefont {Chen}, \citenamefont
  {Demler}, \citenamefont {Lukin},\ and\ \citenamefont {Rey}}]{Gorshkov:11b}%
  \BibitemOpen
  \bibfield  {author} {\bibinfo {author} {\bibfnamefont {A.~V.}\ \bibnamefont
  {Gorshkov}}, \bibinfo {author} {\bibfnamefont {S.~R.}\ \bibnamefont
  {Manmana}}, \bibinfo {author} {\bibfnamefont {G.}~\bibnamefont {Chen}},
  \bibinfo {author} {\bibfnamefont {E.}~\bibnamefont {Demler}}, \bibinfo
  {author} {\bibfnamefont {M.~D.}\ \bibnamefont {Lukin}},\ and\ \bibinfo
  {author} {\bibfnamefont {A.~M.}\ \bibnamefont {Rey}},\ }\bibfield  {title}
  {\bibinfo {title} {Quantum magnetism with polar alkali-metal dimers},\ }\href
  {https://doi.org/10.1103/PhysRevA.84.033619} {\bibfield  {journal} {\bibinfo
  {journal} {Phys. Rev. A}\ }\textbf {\bibinfo {volume} {84}},\ \bibinfo
  {pages} {033619} (\bibinfo {year} {2011})}\BibitemShut {NoStop}%
\bibitem [{\citenamefont {Hazzard}\ \emph {et~al.}(2013)\citenamefont
  {Hazzard}, \citenamefont {Manmana}, \citenamefont {Foss-Feig},\ and\
  \citenamefont {Rey}}]{Hazzard:13}%
  \BibitemOpen
  \bibfield  {author} {\bibinfo {author} {\bibfnamefont {K.~R.~A.}\
  \bibnamefont {Hazzard}}, \bibinfo {author} {\bibfnamefont {S.~R.}\
  \bibnamefont {Manmana}}, \bibinfo {author} {\bibfnamefont {M.}~\bibnamefont
  {Foss-Feig}},\ and\ \bibinfo {author} {\bibfnamefont {A.~M.}\ \bibnamefont
  {Rey}},\ }\bibfield  {title} {\bibinfo {title} {Far-from-equilibrium quantum
  magnetism with ultracold polar molecules},\ }\href
  {https://doi.org/10.1103/PhysRevLett.110.075301} {\bibfield  {journal}
  {\bibinfo  {journal} {Phys. Rev. Lett.}\ }\textbf {\bibinfo {volume} {110}},\
  \bibinfo {pages} {075301} (\bibinfo {year} {2013})}\BibitemShut {NoStop}%
\bibitem [{\citenamefont {Aldegunde}\ \emph {et~al.}(2008)\citenamefont
  {Aldegunde}, \citenamefont {Rivington}, \citenamefont {\ifmmode~\dot{Z}\else
  \.{Z}\fi{}uchowski},\ and\ \citenamefont {Hutson}}]{Aldegunde:08}%
  \BibitemOpen
  \bibfield  {author} {\bibinfo {author} {\bibfnamefont {J.}~\bibnamefont
  {Aldegunde}}, \bibinfo {author} {\bibfnamefont {B.~A.}\ \bibnamefont
  {Rivington}}, \bibinfo {author} {\bibfnamefont {P.~S.}\ \bibnamefont
  {\ifmmode~\dot{Z}\else \.{Z}\fi{}uchowski}},\ and\ \bibinfo {author}
  {\bibfnamefont {J.~M.}\ \bibnamefont {Hutson}},\ }\bibfield  {title}
  {\bibinfo {title} {Hyperfine energy levels of alkali-metal dimers:
  Ground-state polar molecules in electric and magnetic fields},\ }\href
  {https://doi.org/10.1103/PhysRevA.78.033434} {\bibfield  {journal} {\bibinfo
  {journal} {Phys. Rev. A}\ }\textbf {\bibinfo {volume} {78}},\ \bibinfo
  {pages} {033434} (\bibinfo {year} {2008})}\BibitemShut {NoStop}%
\bibitem [{\citenamefont {Brown}\ and\ \citenamefont
  {Carrington}(2003)}]{Brown:03}%
  \BibitemOpen
  \bibfield  {author} {\bibinfo {author} {\bibfnamefont {J.}~\bibnamefont
  {Brown}}\ and\ \bibinfo {author} {\bibfnamefont {A.}~\bibnamefont
  {Carrington}},\ }\href@noop {} {\emph {\bibinfo {title} {{\it Rotational
  Spectroscopy of Diatomic Molecules}}}}\ (\bibinfo  {publisher} {Cambridge
  University Press},\ \bibinfo {year} {2003})\BibitemShut {NoStop}%
\bibitem [{\citenamefont {Fano}(1948)}]{Fano:48}%
  \BibitemOpen
  \bibfield  {author} {\bibinfo {author} {\bibfnamefont {U.}~\bibnamefont
  {Fano}},\ }\bibfield  {title} {\bibinfo {title} {Electric quadrupole coupling
  of the nuclear spin with the rotation of a polar diatomic molecule in an
  external electric field},\ }\href {https://doi.org/10.6028/jres.040.010}
  {\bibfield  {journal} {\bibinfo  {journal} {Journal of Research of the
  National Bureau of Standards}\ }\textbf {\bibinfo {volume} {40}},\ \bibinfo
  {pages} {215} (\bibinfo {year} {1948})}\BibitemShut {NoStop}%
\bibitem [{\citenamefont {Zare}(1988)}]{Zare:88}%
  \BibitemOpen
  \bibfield  {author} {\bibinfo {author} {\bibfnamefont {R.~N.}\ \bibnamefont
  {Zare}},\ }\href@noop {} {\emph {\bibinfo {title} {Angular Momentum}}}\
  (\bibinfo  {publisher} {Wiley},\ \bibinfo {year} {1988})\BibitemShut
  {NoStop}%
\bibitem [{\citenamefont {Aldegunde}\ \emph {et~al.}(2009)\citenamefont
  {Aldegunde}, \citenamefont {Ran},\ and\ \citenamefont
  {Hutson}}]{Aldegunde:09}%
  \BibitemOpen
  \bibfield  {author} {\bibinfo {author} {\bibfnamefont {J.}~\bibnamefont
  {Aldegunde}}, \bibinfo {author} {\bibfnamefont {H.}~\bibnamefont {Ran}},\
  and\ \bibinfo {author} {\bibfnamefont {J.~M.}\ \bibnamefont {Hutson}},\
  }\bibfield  {title} {\bibinfo {title} {Manipulating ultracold polar molecules
  with microwave radiation: The influence of hyperfine structure},\ }\href
  {https://doi.org/10.1103/PhysRevA.80.043410} {\bibfield  {journal} {\bibinfo
  {journal} {Phys. Rev. A}\ }\textbf {\bibinfo {volume} {80}},\ \bibinfo
  {pages} {043410} (\bibinfo {year} {2009})}\BibitemShut {NoStop}%
\bibitem [{\citenamefont {Nordholm}\ and\ \citenamefont
  {Rice}(1974)}]{Nordholm:74}%
  \BibitemOpen
  \bibfield  {author} {\bibinfo {author} {\bibfnamefont {K.~S.~J.}\
  \bibnamefont {Nordholm}}\ and\ \bibinfo {author} {\bibfnamefont {S.~A.}\
  \bibnamefont {Rice}},\ }\bibfield  {title} {\bibinfo {title} {Quantum
  ergodicity and vibrational relaxation in isolated molecules},\ }\href
  {https://doi.org/10.1063/1.1681624} {\bibfield  {journal} {\bibinfo
  {journal} {J. Chem. Phys.}\ }\textbf {\bibinfo {volume} {61}},\ \bibinfo
  {pages} {203} (\bibinfo {year} {1974})}\BibitemShut {NoStop}%
\bibitem [{\citenamefont {Pittman}\ and\ \citenamefont
  {Heller}(2015)}]{Pittman:15}%
  \BibitemOpen
  \bibfield  {author} {\bibinfo {author} {\bibfnamefont {S.~M.}\ \bibnamefont
  {Pittman}}\ and\ \bibinfo {author} {\bibfnamefont {E.~J.}\ \bibnamefont
  {Heller}},\ }\bibfield  {title} {\bibinfo {title} {The degree of ergodicity
  of ortho- and para-aminobenzonitrile in an electric field},\ }\href
  {https://doi.org/10.1021/acs.jpca.5b04838} {\bibfield  {journal} {\bibinfo
  {journal} {J. Phys. Chem. A}\ }\textbf {\bibinfo {volume} {119}},\ \bibinfo
  {pages} {10563} (\bibinfo {year} {2015})}\BibitemShut {NoStop}%
\bibitem [{\citenamefont {Uzer}\ and\ \citenamefont {Miller}(1991)}]{Uzer:91}%
  \BibitemOpen
  \bibfield  {author} {\bibinfo {author} {\bibfnamefont {T.}~\bibnamefont
  {Uzer}}\ and\ \bibinfo {author} {\bibfnamefont {W.~H.}\ \bibnamefont
  {Miller}},\ }\bibfield  {title} {\bibinfo {title} {Theories of intramolecular
  vibrational energy transfer},\ }\href
  {https://doi.org/https://doi.org/10.1016/0370-1573(91)90140-H} {\bibfield
  {journal} {\bibinfo  {journal} {Phys. Rep.}\ }\textbf {\bibinfo {volume}
  {199}},\ \bibinfo {pages} {73} (\bibinfo {year} {1991})}\BibitemShut
  {NoStop}%
\bibitem [{\citenamefont {Bigwood}\ \emph {et~al.}(1998)\citenamefont
  {Bigwood}, \citenamefont {Gruebele}, \citenamefont {Leitner},\ and\
  \citenamefont {Wolynes}}]{Bigwood:98}%
  \BibitemOpen
  \bibfield  {author} {\bibinfo {author} {\bibfnamefont {R.}~\bibnamefont
  {Bigwood}}, \bibinfo {author} {\bibfnamefont {M.}~\bibnamefont {Gruebele}},
  \bibinfo {author} {\bibfnamefont {D.~M.}\ \bibnamefont {Leitner}},\ and\
  \bibinfo {author} {\bibfnamefont {P.~G.}\ \bibnamefont {Wolynes}},\
  }\bibfield  {title} {\bibinfo {title} {The vibrational energy flow transition
  in organic molecules: Theory meets experiment},\ }\href
  {https://doi.org/10.1073/pnas.95.11.5960} {\bibfield  {journal} {\bibinfo
  {journal} {Proc. Natl. Acad. Sci. USA}\ }\textbf {\bibinfo {volume} {95}},\
  \bibinfo {pages} {5960} (\bibinfo {year} {1998})}\BibitemShut {NoStop}%
\bibitem [{\citenamefont {Nesbitt}\ and\ \citenamefont
  {Field}(1996)}]{Nesbitt:96}%
  \BibitemOpen
  \bibfield  {author} {\bibinfo {author} {\bibfnamefont {D.~J.}\ \bibnamefont
  {Nesbitt}}\ and\ \bibinfo {author} {\bibfnamefont {R.~W.}\ \bibnamefont
  {Field}},\ }\bibfield  {title} {\bibinfo {title} {Vibrational energy flow in
  highly excited molecules: Role of intramolecular vibrational
  redistribution},\ }\href {https://doi.org/10.1021/jp960698w} {\bibfield
  {journal} {\bibinfo  {journal} {J. Phys. Chem.}\ }\textbf {\bibinfo {volume}
  {100}},\ \bibinfo {pages} {12735} (\bibinfo {year} {1996})}\BibitemShut
  {NoStop}%
\bibitem [{\citenamefont {Leitner}(2015)}]{Leitner:15}%
  \BibitemOpen
  \bibfield  {author} {\bibinfo {author} {\bibfnamefont {D.~M.}\ \bibnamefont
  {Leitner}},\ }\bibfield  {title} {\bibinfo {title} {Quantum ergodicity and
  energy flow in molecules},\ }\href
  {https://doi.org/10.1080/00018732.2015.1109817} {\bibfield  {journal}
  {\bibinfo  {journal} {Adv. Phys.}\ }\textbf {\bibinfo {volume} {64}},\
  \bibinfo {pages} {445} (\bibinfo {year} {2015})}\BibitemShut {NoStop}%
\bibitem [{\citenamefont {Liu}\ \emph {et~al.}(2023)\citenamefont {Liu},
  \citenamefont {Rosenberg}, \citenamefont {Changala}, \citenamefont {Crowley},
  \citenamefont {Nesbitt}, \citenamefont {Yao}, \citenamefont {Tscherbul},\
  and\ \citenamefont {Ye}}]{Liu:23}%
  \BibitemOpen
  \bibfield  {author} {\bibinfo {author} {\bibfnamefont {L.~R.}\ \bibnamefont
  {Liu}}, \bibinfo {author} {\bibfnamefont {D.}~\bibnamefont {Rosenberg}},
  \bibinfo {author} {\bibfnamefont {P.~B.}\ \bibnamefont {Changala}}, \bibinfo
  {author} {\bibfnamefont {P.~J.~D.}\ \bibnamefont {Crowley}}, \bibinfo
  {author} {\bibfnamefont {D.~J.}\ \bibnamefont {Nesbitt}}, \bibinfo {author}
  {\bibfnamefont {N.~Y.}\ \bibnamefont {Yao}}, \bibinfo {author} {\bibfnamefont
  {T.~V.}\ \bibnamefont {Tscherbul}},\ and\ \bibinfo {author} {\bibfnamefont
  {J.}~\bibnamefont {Ye}},\ }\bibfield  {title} {\bibinfo {title} {Ergodicity
  breaking in rapidly rotating {C$_{60}$} fullerenes},\ }\href
  {https://doi.org/10.1126/science.adi6354} {\bibfield  {journal} {\bibinfo
  {journal} {Science}\ }\textbf {\bibinfo {volume} {381}},\ \bibinfo {pages}
  {778} (\bibinfo {year} {2023})}\BibitemShut {NoStop}%
\bibitem [{\citenamefont {Kotochigova}\ and\ \citenamefont
  {DeMille}(2010)}]{Kotochigova:10b}%
  \BibitemOpen
  \bibfield  {author} {\bibinfo {author} {\bibfnamefont {S.}~\bibnamefont
  {Kotochigova}}\ and\ \bibinfo {author} {\bibfnamefont {D.}~\bibnamefont
  {DeMille}},\ }\bibfield  {title} {\bibinfo {title} {Electric-field-dependent
  dynamic polarizability and state-insensitive conditions for optical trapping
  of diatomic polar molecules},\ }\href
  {https://doi.org/10.1103/PhysRevA.82.063421} {\bibfield  {journal} {\bibinfo
  {journal} {Phys. Rev. A}\ }\textbf {\bibinfo {volume} {82}},\ \bibinfo
  {pages} {063421} (\bibinfo {year} {2010})}\BibitemShut {NoStop}%
\bibitem [{\citenamefont {Neyenhuis}\ \emph {et~al.}(2012)\citenamefont
  {Neyenhuis}, \citenamefont {Yan}, \citenamefont {Moses}, \citenamefont
  {Covey}, \citenamefont {Chotia}, \citenamefont {Petrov}, \citenamefont
  {Kotochigova}, \citenamefont {Ye},\ and\ \citenamefont {Jin}}]{Neyenhuis:12}%
  \BibitemOpen
  \bibfield  {author} {\bibinfo {author} {\bibfnamefont {B.}~\bibnamefont
  {Neyenhuis}}, \bibinfo {author} {\bibfnamefont {B.}~\bibnamefont {Yan}},
  \bibinfo {author} {\bibfnamefont {S.~A.}\ \bibnamefont {Moses}}, \bibinfo
  {author} {\bibfnamefont {J.~P.}\ \bibnamefont {Covey}}, \bibinfo {author}
  {\bibfnamefont {A.}~\bibnamefont {Chotia}}, \bibinfo {author} {\bibfnamefont
  {A.}~\bibnamefont {Petrov}}, \bibinfo {author} {\bibfnamefont
  {S.}~\bibnamefont {Kotochigova}}, \bibinfo {author} {\bibfnamefont
  {J.}~\bibnamefont {Ye}},\ and\ \bibinfo {author} {\bibfnamefont {D.~S.}\
  \bibnamefont {Jin}},\ }\bibfield  {title} {\bibinfo {title} {{Anisotropic
  Polarizability of Ultracold Polar $^{40}\mathrm{K}^{87}\mathrm{Rb}$
  Molecules}},\ }\href {https://doi.org/10.1103/PhysRevLett.109.230403}
  {\bibfield  {journal} {\bibinfo  {journal} {Phys. Rev. Lett.}\ }\textbf
  {\bibinfo {volume} {109}},\ \bibinfo {pages} {230403} (\bibinfo {year}
  {2012})}\BibitemShut {NoStop}%
\bibitem [{\citenamefont {Blackmore}\ \emph {et~al.}(2018)\citenamefont
  {Blackmore}, \citenamefont {Caldwell}, \citenamefont {Gregory}, \citenamefont
  {Bridge}, \citenamefont {Sawant}, \citenamefont {Aldegunde}, \citenamefont
  {Mur-Petit}, \citenamefont {Jaksch}, \citenamefont {Hutson}, \citenamefont
  {Sauer}, \citenamefont {Tarbutt},\ and\ \citenamefont
  {Cornish}}]{Blackmore:18}%
  \BibitemOpen
  \bibfield  {author} {\bibinfo {author} {\bibfnamefont {J.~A.}\ \bibnamefont
  {Blackmore}}, \bibinfo {author} {\bibfnamefont {L.}~\bibnamefont {Caldwell}},
  \bibinfo {author} {\bibfnamefont {P.~D.}\ \bibnamefont {Gregory}}, \bibinfo
  {author} {\bibfnamefont {E.~M.}\ \bibnamefont {Bridge}}, \bibinfo {author}
  {\bibfnamefont {R.}~\bibnamefont {Sawant}}, \bibinfo {author} {\bibfnamefont
  {J.}~\bibnamefont {Aldegunde}}, \bibinfo {author} {\bibfnamefont
  {J.}~\bibnamefont {Mur-Petit}}, \bibinfo {author} {\bibfnamefont
  {D.}~\bibnamefont {Jaksch}}, \bibinfo {author} {\bibfnamefont {J.~M.}\
  \bibnamefont {Hutson}}, \bibinfo {author} {\bibfnamefont {B.~E.}\
  \bibnamefont {Sauer}}, \bibinfo {author} {\bibfnamefont {M.~R.}\ \bibnamefont
  {Tarbutt}},\ and\ \bibinfo {author} {\bibfnamefont {S.~L.}\ \bibnamefont
  {Cornish}},\ }\bibfield  {title} {\bibinfo {title} {Ultracold molecules for
  quantum simulation: rotational coherences in {CaF and RbCs}},\ }\href
  {https://doi.org/10.1088/2058-9565/aaee35} {\bibfield  {journal} {\bibinfo
  {journal} {Quantum Sci. Technol.}\ }\textbf {\bibinfo {volume} {4}},\
  \bibinfo {pages} {014010} (\bibinfo {year} {2018})}\BibitemShut {NoStop}%
\bibitem [{\citenamefont {See\ss{}elberg}\ \emph {et~al.}(2018)\citenamefont
  {See\ss{}elberg}, \citenamefont {Luo}, \citenamefont {Li}, \citenamefont
  {Bause}, \citenamefont {Kotochigova}, \citenamefont {Bloch},\ and\
  \citenamefont {Gohle}}]{Sesselberg:18}%
  \BibitemOpen
  \bibfield  {author} {\bibinfo {author} {\bibfnamefont {F.}~\bibnamefont
  {See\ss{}elberg}}, \bibinfo {author} {\bibfnamefont {X.-Y.}\ \bibnamefont
  {Luo}}, \bibinfo {author} {\bibfnamefont {M.}~\bibnamefont {Li}}, \bibinfo
  {author} {\bibfnamefont {R.}~\bibnamefont {Bause}}, \bibinfo {author}
  {\bibfnamefont {S.}~\bibnamefont {Kotochigova}}, \bibinfo {author}
  {\bibfnamefont {I.}~\bibnamefont {Bloch}},\ and\ \bibinfo {author}
  {\bibfnamefont {C.}~\bibnamefont {Gohle}},\ }\bibfield  {title} {\bibinfo
  {title} {Extending rotational coherence of interacting polar molecules in a
  spin-decoupled magic trap},\ }\href
  {https://doi.org/10.1103/PhysRevLett.121.253401} {\bibfield  {journal}
  {\bibinfo  {journal} {Phys. Rev. Lett.}\ }\textbf {\bibinfo {volume} {121}},\
  \bibinfo {pages} {253401} (\bibinfo {year} {2018})}\BibitemShut {NoStop}%
\bibitem [{\citenamefont {Gregory}\ \emph {et~al.}(2021)\citenamefont
  {Gregory}, \citenamefont {Blackmore}, \citenamefont {Bromley}, \citenamefont
  {Hutson},\ and\ \citenamefont {Cornish}}]{Gregory:21}%
  \BibitemOpen
  \bibfield  {author} {\bibinfo {author} {\bibfnamefont {P.~D.}\ \bibnamefont
  {Gregory}}, \bibinfo {author} {\bibfnamefont {J.~A.}\ \bibnamefont
  {Blackmore}}, \bibinfo {author} {\bibfnamefont {S.~L.}\ \bibnamefont
  {Bromley}}, \bibinfo {author} {\bibfnamefont {J.~M.}\ \bibnamefont
  {Hutson}},\ and\ \bibinfo {author} {\bibfnamefont {S.~L.}\ \bibnamefont
  {Cornish}},\ }\bibfield  {title} {\bibinfo {title} {Robust storage qubits in
  ultracold polar molecules},\ }\href
  {https://doi.org/10.1038/s41567-021-01328-7} {\bibfield  {journal} {\bibinfo
  {journal} {Nat. Phys.}\ }\textbf {\bibinfo {volume} {17}},\ \bibinfo {pages}
  {1149} (\bibinfo {year} {2021})}\BibitemShut {NoStop}%
\bibitem [{\citenamefont {Lin}\ \emph {et~al.}(2022)\citenamefont {Lin},
  \citenamefont {He}, \citenamefont {Jin}, \citenamefont {Chen},\ and\
  \citenamefont {Wang}}]{Lin:22}%
  \BibitemOpen
  \bibfield  {author} {\bibinfo {author} {\bibfnamefont {J.}~\bibnamefont
  {Lin}}, \bibinfo {author} {\bibfnamefont {J.}~\bibnamefont {He}}, \bibinfo
  {author} {\bibfnamefont {M.}~\bibnamefont {Jin}}, \bibinfo {author}
  {\bibfnamefont {G.}~\bibnamefont {Chen}},\ and\ \bibinfo {author}
  {\bibfnamefont {D.}~\bibnamefont {Wang}},\ }\bibfield  {title} {\bibinfo
  {title} {Seconds-scale coherence on nuclear spin transitions of ultracold
  polar molecules in 3d optical lattices},\ }\href
  {https://doi.org/10.1103/PhysRevLett.128.223201} {\bibfield  {journal}
  {\bibinfo  {journal} {Phys. Rev. Lett.}\ }\textbf {\bibinfo {volume} {128}},\
  \bibinfo {pages} {223201} (\bibinfo {year} {2022})}\BibitemShut {NoStop}%
\bibitem [{\citenamefont {Wall}\ \emph {et~al.}(2015)\citenamefont {Wall},
  \citenamefont {Hazzard},\ and\ \citenamefont {Rey}}]{Wall:15c}%
  \BibitemOpen
  \bibfield  {author} {\bibinfo {author} {\bibfnamefont {M.~L.}\ \bibnamefont
  {Wall}}, \bibinfo {author} {\bibfnamefont {K.~R.~A.}\ \bibnamefont
  {Hazzard}},\ and\ \bibinfo {author} {\bibfnamefont {A.~M.}\ \bibnamefont
  {Rey}},\ }\bibinfo {title} {Quantum magnetism with ultracold molecules},\ in\
  \href {https://doi.org/https://doi.org/10.1142/9789814678704_0001} {\emph
  {\bibinfo {booktitle} {From Atomic to Mesoscale. The Role of Quantum
  Coherence in Systems of Various Complexities}}}\ (\bibinfo  {publisher}
  {{World Scientific}},\ \bibinfo {year} {2015})\ pp.\ \bibinfo {pages}
  {3--37}\BibitemShut {NoStop}%
\bibitem [{\citenamefont {Perlin}\ \emph {et~al.}(2020)\citenamefont {Perlin},
  \citenamefont {Qu},\ and\ \citenamefont {Rey}}]{Perlin:20}%
  \BibitemOpen
  \bibfield  {author} {\bibinfo {author} {\bibfnamefont {M.~A.}\ \bibnamefont
  {Perlin}}, \bibinfo {author} {\bibfnamefont {C.}~\bibnamefont {Qu}},\ and\
  \bibinfo {author} {\bibfnamefont {A.~M.}\ \bibnamefont {Rey}},\ }\bibfield
  {title} {\bibinfo {title} {Spin squeezing with short-range spin-exchange
  interactions},\ }\href {https://doi.org/10.1103/PhysRevLett.125.223401}
  {\bibfield  {journal} {\bibinfo  {journal} {Phys. Rev. Lett.}\ }\textbf
  {\bibinfo {volume} {125}},\ \bibinfo {pages} {223401} (\bibinfo {year}
  {2020})}\BibitemShut {NoStop}%
\bibitem [{\citenamefont {Chang}\ \emph {et~al.}(2004)\citenamefont {Chang},
  \citenamefont {Ye},\ and\ \citenamefont {Lukin}}]{Chang:04}%
  \BibitemOpen
  \bibfield  {author} {\bibinfo {author} {\bibfnamefont {D.~E.}\ \bibnamefont
  {Chang}}, \bibinfo {author} {\bibfnamefont {J.}~\bibnamefont {Ye}},\ and\
  \bibinfo {author} {\bibfnamefont {M.~D.}\ \bibnamefont {Lukin}},\ }\bibfield
  {title} {\bibinfo {title} {Controlling dipole-dipole frequency shifts in a
  lattice-based optical atomic clock},\ }\href
  {https://doi.org/10.1103/PhysRevA.69.023810} {\bibfield  {journal} {\bibinfo
  {journal} {Phys. Rev. A}\ }\textbf {\bibinfo {volume} {69}},\ \bibinfo
  {pages} {023810} (\bibinfo {year} {2004})}\BibitemShut {NoStop}%
\bibitem [{\citenamefont {Bai}\ \emph {et~al.}(2023)\citenamefont {Bai},
  \citenamefont {Wen}, \citenamefont {Peng}, \citenamefont {Liu},\ and\
  \citenamefont {Luo}}]{Bai2023}%
  \BibitemOpen
  \bibfield  {author} {\bibinfo {author} {\bibfnamefont {X.}~\bibnamefont
  {Bai}}, \bibinfo {author} {\bibfnamefont {K.}~\bibnamefont {Wen}}, \bibinfo
  {author} {\bibfnamefont {D.}~\bibnamefont {Peng}}, \bibinfo {author}
  {\bibfnamefont {S.}~\bibnamefont {Liu}},\ and\ \bibinfo {author}
  {\bibfnamefont {L.}~\bibnamefont {Luo}},\ }\bibfield  {title} {\bibinfo
  {title} {Atomic magnetometers and their application in industry},\ }\bibfield
   {journal} {\bibinfo  {journal} {Front. Phys.}\ }\textbf {\bibinfo {volume}
  {11}},\ \href {https://doi.org/10.3389/fphy.2023.1212368}
  {10.3389/fphy.2023.1212368} (\bibinfo {year} {2023})\BibitemShut {NoStop}%
\bibitem [{\citenamefont {Wang}\ \emph {et~al.}(2020)\citenamefont {Wang},
  \citenamefont {Hu}, \citenamefont {Sanders},\ and\ \citenamefont
  {Kais}}]{Wang:20}%
  \BibitemOpen
  \bibfield  {author} {\bibinfo {author} {\bibfnamefont {Y.}~\bibnamefont
  {Wang}}, \bibinfo {author} {\bibfnamefont {Z.}~\bibnamefont {Hu}}, \bibinfo
  {author} {\bibfnamefont {B.~C.}\ \bibnamefont {Sanders}},\ and\ \bibinfo
  {author} {\bibfnamefont {S.}~\bibnamefont {Kais}},\ }\bibfield  {title}
  {\bibinfo {title} {Qudits and high-dimensional quantum computing},\ }\href
  {https://doi.org/10.3389/fphy.2020.589504} {\bibfield  {journal} {\bibinfo
  {journal} {Front. Phys.}\ }\textbf {\bibinfo {volume} {8}},\ \bibinfo {pages}
  {589504} (\bibinfo {year} {2020})}\BibitemShut {NoStop}%
\bibitem [{\citenamefont {Sawant}\ \emph {et~al.}(2020)\citenamefont {Sawant},
  \citenamefont {Blackmore}, \citenamefont {Gregory}, \citenamefont
  {Mur-Petit}, \citenamefont {Jaksch}, \citenamefont {Aldegunde}, \citenamefont
  {Hutson}, \citenamefont {Tarbutt},\ and\ \citenamefont
  {Cornish}}]{Sawant:20}%
  \BibitemOpen
  \bibfield  {author} {\bibinfo {author} {\bibfnamefont {R.}~\bibnamefont
  {Sawant}}, \bibinfo {author} {\bibfnamefont {J.~A.}\ \bibnamefont
  {Blackmore}}, \bibinfo {author} {\bibfnamefont {P.~D.}\ \bibnamefont
  {Gregory}}, \bibinfo {author} {\bibfnamefont {J.}~\bibnamefont {Mur-Petit}},
  \bibinfo {author} {\bibfnamefont {D.}~\bibnamefont {Jaksch}}, \bibinfo
  {author} {\bibfnamefont {J.}~\bibnamefont {Aldegunde}}, \bibinfo {author}
  {\bibfnamefont {J.~M.}\ \bibnamefont {Hutson}}, \bibinfo {author}
  {\bibfnamefont {M.~R.}\ \bibnamefont {Tarbutt}},\ and\ \bibinfo {author}
  {\bibfnamefont {S.~L.}\ \bibnamefont {Cornish}},\ }\bibfield  {title}
  {\bibinfo {title} {Ultracold polar molecules as qudits},\ }\href
  {https://doi.org/10.1088/1367-2630/ab60f4} {\bibfield  {journal} {\bibinfo
  {journal} {New J. Phys.}\ }\textbf {\bibinfo {volume} {22}},\ \bibinfo
  {pages} {013027} (\bibinfo {year} {2020})}\BibitemShut {NoStop}%
\bibitem [{\citenamefont {Gonz\'alez-Cuadra}\ \emph {et~al.}(2022)\citenamefont
  {Gonz\'alez-Cuadra}, \citenamefont {Zache}, \citenamefont {Carrasco},
  \citenamefont {Kraus},\ and\ \citenamefont {Zoller}}]{GonzalezCuadra:22}%
  \BibitemOpen
  \bibfield  {author} {\bibinfo {author} {\bibfnamefont {D.}~\bibnamefont
  {Gonz\'alez-Cuadra}}, \bibinfo {author} {\bibfnamefont {T.~V.}\ \bibnamefont
  {Zache}}, \bibinfo {author} {\bibfnamefont {J.}~\bibnamefont {Carrasco}},
  \bibinfo {author} {\bibfnamefont {B.}~\bibnamefont {Kraus}},\ and\ \bibinfo
  {author} {\bibfnamefont {P.}~\bibnamefont {Zoller}},\ }\bibfield  {title}
  {\bibinfo {title} {Hardware efficient quantum simulation of non-abelian gauge
  theories with qudits on {Rydberg} platforms},\ }\href
  {https://doi.org/10.1103/PhysRevLett.129.160501} {\bibfield  {journal}
  {\bibinfo  {journal} {Phys. Rev. Lett.}\ }\textbf {\bibinfo {volume} {129}},\
  \bibinfo {pages} {160501} (\bibinfo {year} {2022})}\BibitemShut {NoStop}%
\bibitem [{\citenamefont {Di}\ \emph {et~al.}(2010)\citenamefont {Di},
  \citenamefont {Wang},\ and\ \citenamefont {Wei}}]{Di:10}%
  \BibitemOpen
  \bibfield  {author} {\bibinfo {author} {\bibfnamefont {Y.}~\bibnamefont
  {Di}}, \bibinfo {author} {\bibfnamefont {Y.}~\bibnamefont {Wang}},\ and\
  \bibinfo {author} {\bibfnamefont {H.}~\bibnamefont {Wei}},\ }\bibfield
  {title} {\bibinfo {title} {Dipole–quadrupole decomposition of two coupled
  spin 1 systems},\ }\href {https://doi.org/10.1088/1751-8113/43/6/065303}
  {\bibfield  {journal} {\bibinfo  {journal} {J. Phys. A}\ }\textbf {\bibinfo
  {volume} {43}},\ \bibinfo {pages} {065303} (\bibinfo {year}
  {2010})}\BibitemShut {NoStop}%
\bibitem [{\citenamefont {Hamley}\ \emph
  {et~al.}(2012{\natexlab{a}})\citenamefont {Hamley}, \citenamefont {Gerving},
  \citenamefont {Hoang}, \citenamefont {Bookjans},\ and\ \citenamefont
  {Chapman}}]{Hamley:12}%
  \BibitemOpen
  \bibfield  {author} {\bibinfo {author} {\bibfnamefont {C.~D.}\ \bibnamefont
  {Hamley}}, \bibinfo {author} {\bibfnamefont {C.~S.}\ \bibnamefont {Gerving}},
  \bibinfo {author} {\bibfnamefont {T.~M.}\ \bibnamefont {Hoang}}, \bibinfo
  {author} {\bibfnamefont {E.~M.}\ \bibnamefont {Bookjans}},\ and\ \bibinfo
  {author} {\bibfnamefont {M.~S.}\ \bibnamefont {Chapman}},\ }\bibfield
  {title} {\bibinfo {title} {Spin-nematic squeezed vacuum in a quantum gas},\
  }\href {https://doi.org/10.1038/nphys2245} {\bibfield  {journal} {\bibinfo
  {journal} {Nat. Phys.}\ }\textbf {\bibinfo {volume} {8}},\ \bibinfo {pages}
  {305} (\bibinfo {year} {2012}{\natexlab{a}})}\BibitemShut {NoStop}%
\bibitem [{\citenamefont {Lindon}\ \emph {et~al.}(2023)\citenamefont {Lindon},
  \citenamefont {Tashchilina}, \citenamefont {Cooke},\ and\ \citenamefont
  {LeBlanc}}]{LeBlanc:23}%
  \BibitemOpen
  \bibfield  {author} {\bibinfo {author} {\bibfnamefont {J.}~\bibnamefont
  {Lindon}}, \bibinfo {author} {\bibfnamefont {A.}~\bibnamefont {Tashchilina}},
  \bibinfo {author} {\bibfnamefont {L.~W.}\ \bibnamefont {Cooke}},\ and\
  \bibinfo {author} {\bibfnamefont {L.~J.}\ \bibnamefont {LeBlanc}},\
  }\bibfield  {title} {\bibinfo {title} {Complete unitary qutrit control in
  ultracold atoms},\ }\href {https://doi.org/10.1103/PhysRevApplied.19.034089}
  {\bibfield  {journal} {\bibinfo  {journal} {Phys. Rev. Appl.}\ }\textbf
  {\bibinfo {volume} {19}},\ \bibinfo {pages} {034089} (\bibinfo {year}
  {2023})}\BibitemShut {NoStop}%
\bibitem [{\citenamefont {Barry}\ \emph {et~al.}(2014)\citenamefont {Barry},
  \citenamefont {McCarron}, \citenamefont {Norrgard}, \citenamefont
  {Steinecker},\ and\ \citenamefont {DeMille}}]{Barry:14}%
  \BibitemOpen
  \bibfield  {author} {\bibinfo {author} {\bibfnamefont {J.~F.}\ \bibnamefont
  {Barry}}, \bibinfo {author} {\bibfnamefont {D.~J.}\ \bibnamefont {McCarron}},
  \bibinfo {author} {\bibfnamefont {E.~B.}\ \bibnamefont {Norrgard}}, \bibinfo
  {author} {\bibfnamefont {M.~H.}\ \bibnamefont {Steinecker}},\ and\ \bibinfo
  {author} {\bibfnamefont {D.}~\bibnamefont {DeMille}},\ }\bibfield  {title}
  {\bibinfo {title} {Magneto-optical trapping of a diatomic molecule},\
  }\href@noop {} {\bibfield  {journal} {\bibinfo  {journal} {Nature (London)}\
  }\textbf {\bibinfo {volume} {512}},\ \bibinfo {pages} {286} (\bibinfo {year}
  {2014})}\BibitemShut {NoStop}%
\bibitem [{\citenamefont {Collopy}\ \emph {et~al.}(2018)\citenamefont
  {Collopy}, \citenamefont {Ding}, \citenamefont {Wu}, \citenamefont
  {Finneran}, \citenamefont {Anderegg}, \citenamefont {Augenbraun},
  \citenamefont {Doyle},\ and\ \citenamefont {Ye}}]{Collopy:18}%
  \BibitemOpen
  \bibfield  {author} {\bibinfo {author} {\bibfnamefont {A.~L.}\ \bibnamefont
  {Collopy}}, \bibinfo {author} {\bibfnamefont {S.}~\bibnamefont {Ding}},
  \bibinfo {author} {\bibfnamefont {Y.}~\bibnamefont {Wu}}, \bibinfo {author}
  {\bibfnamefont {I.~A.}\ \bibnamefont {Finneran}}, \bibinfo {author}
  {\bibfnamefont {L.}~\bibnamefont {Anderegg}}, \bibinfo {author}
  {\bibfnamefont {B.~L.}\ \bibnamefont {Augenbraun}}, \bibinfo {author}
  {\bibfnamefont {J.~M.}\ \bibnamefont {Doyle}},\ and\ \bibinfo {author}
  {\bibfnamefont {J.}~\bibnamefont {Ye}},\ }\bibfield  {title} {\bibinfo
  {title} {{3D} magneto-optical trap of yttrium monoxide},\ }\href
  {https://doi.org/10.1103/PhysRevLett.121.213201} {\bibfield  {journal}
  {\bibinfo  {journal} {Phys. Rev. Lett.}\ }\textbf {\bibinfo {volume} {121}},\
  \bibinfo {pages} {213201} (\bibinfo {year} {2018})}\BibitemShut {NoStop}%
\bibitem [{\citenamefont {McCarron}\ \emph {et~al.}(2018)\citenamefont
  {McCarron}, \citenamefont {Steinecker}, \citenamefont {Zhu},\ and\
  \citenamefont {DeMille}}]{McCarron:18}%
  \BibitemOpen
  \bibfield  {author} {\bibinfo {author} {\bibfnamefont {D.~J.}\ \bibnamefont
  {McCarron}}, \bibinfo {author} {\bibfnamefont {M.~H.}\ \bibnamefont
  {Steinecker}}, \bibinfo {author} {\bibfnamefont {Y.}~\bibnamefont {Zhu}},\
  and\ \bibinfo {author} {\bibfnamefont {D.}~\bibnamefont {DeMille}},\
  }\bibfield  {title} {\bibinfo {title} {Magnetic trapping of an ultracold gas
  of polar molecules},\ }\href {https://doi.org/10.1103/PhysRevLett.121.013202}
  {\bibfield  {journal} {\bibinfo  {journal} {Phys. Rev. Lett.}\ }\textbf
  {\bibinfo {volume} {121}},\ \bibinfo {pages} {013202} (\bibinfo {year}
  {2018})}\BibitemShut {NoStop}%
\bibitem [{\citenamefont {Changala}\ \emph {et~al.}(2019)\citenamefont
  {Changala}, \citenamefont {Weichman}, \citenamefont {Lee}, \citenamefont
  {Fermann},\ and\ \citenamefont {Ye}}]{Changala:19}%
  \BibitemOpen
  \bibfield  {author} {\bibinfo {author} {\bibfnamefont {P.~B.}\ \bibnamefont
  {Changala}}, \bibinfo {author} {\bibfnamefont {M.~L.}\ \bibnamefont
  {Weichman}}, \bibinfo {author} {\bibfnamefont {K.~F.}\ \bibnamefont {Lee}},
  \bibinfo {author} {\bibfnamefont {M.~E.}\ \bibnamefont {Fermann}},\ and\
  \bibinfo {author} {\bibfnamefont {J.}~\bibnamefont {Ye}},\ }\bibfield
  {title} {\bibinfo {title} {Rovibrational quantum state resolution of the
  {C$_{60}$} fullerene},\ }\href {https://doi.org/10.1126/science.aav2616}
  {\bibfield  {journal} {\bibinfo  {journal} {Science}\ }\textbf {\bibinfo
  {volume} {363}},\ \bibinfo {pages} {49} (\bibinfo {year} {2019})}\BibitemShut
  {NoStop}%
\bibitem [{\citenamefont {Anderegg}\ \emph {et~al.}(2018)\citenamefont
  {Anderegg}, \citenamefont {Augenbraun}, \citenamefont {Bao}, \citenamefont
  {Burchesky}, \citenamefont {Cheuk}, \citenamefont {Ketterle},\ and\
  \citenamefont {Doyle}}]{Anderegg:18}%
  \BibitemOpen
  \bibfield  {author} {\bibinfo {author} {\bibfnamefont {L.}~\bibnamefont
  {Anderegg}}, \bibinfo {author} {\bibfnamefont {B.~L.}\ \bibnamefont
  {Augenbraun}}, \bibinfo {author} {\bibfnamefont {Y.}~\bibnamefont {Bao}},
  \bibinfo {author} {\bibfnamefont {S.}~\bibnamefont {Burchesky}}, \bibinfo
  {author} {\bibfnamefont {L.~W.}\ \bibnamefont {Cheuk}}, \bibinfo {author}
  {\bibfnamefont {W.}~\bibnamefont {Ketterle}},\ and\ \bibinfo {author}
  {\bibfnamefont {J.~M.}\ \bibnamefont {Doyle}},\ }\bibfield  {title} {\bibinfo
  {title} {Laser cooling of optically trapped molecules},\ }\href
  {https://doi.org/10.1038/s41567-018-0191-z} {\bibfield  {journal} {\bibinfo
  {journal} {Nat. Phys.}\ }\textbf {\bibinfo {volume} {14}},\ \bibinfo {pages}
  {890} (\bibinfo {year} {2018})}\BibitemShut {NoStop}%
\bibitem [{\citenamefont {Burau}\ \emph {et~al.}(2023)\citenamefont {Burau},
  \citenamefont {Aggarwal}, \citenamefont {Mehling},\ and\ \citenamefont
  {Ye}}]{Burau:23}%
  \BibitemOpen
  \bibfield  {author} {\bibinfo {author} {\bibfnamefont {J.~J.}\ \bibnamefont
  {Burau}}, \bibinfo {author} {\bibfnamefont {P.}~\bibnamefont {Aggarwal}},
  \bibinfo {author} {\bibfnamefont {K.}~\bibnamefont {Mehling}},\ and\ \bibinfo
  {author} {\bibfnamefont {J.}~\bibnamefont {Ye}},\ }\bibfield  {title}
  {\bibinfo {title} {Blue-detuned magneto-optical trap of molecules},\ }\href
  {https://doi.org/10.1103/PhysRevLett.130.193401} {\bibfield  {journal}
  {\bibinfo  {journal} {Phys. Rev. Lett.}\ }\textbf {\bibinfo {volume} {130}},\
  \bibinfo {pages} {193401} (\bibinfo {year} {2023})}\BibitemShut {NoStop}%
\bibitem [{\citenamefont {Prehn}\ \emph {et~al.}(2016)\citenamefont {Prehn},
  \citenamefont {Ibr\"ugger}, \citenamefont {Gl\"ockner}, \citenamefont
  {Rempe},\ and\ \citenamefont {Zeppenfeld}}]{Prehn:16}%
  \BibitemOpen
  \bibfield  {author} {\bibinfo {author} {\bibfnamefont {A.}~\bibnamefont
  {Prehn}}, \bibinfo {author} {\bibfnamefont {M.}~\bibnamefont {Ibr\"ugger}},
  \bibinfo {author} {\bibfnamefont {R.}~\bibnamefont {Gl\"ockner}}, \bibinfo
  {author} {\bibfnamefont {G.}~\bibnamefont {Rempe}},\ and\ \bibinfo {author}
  {\bibfnamefont {M.}~\bibnamefont {Zeppenfeld}},\ }\bibfield  {title}
  {\bibinfo {title} {Optoelectrical cooling of polar molecules to
  submillikelvin temperatures},\ }\href
  {https://doi.org/10.1103/PhysRevLett.116.063005} {\bibfield  {journal}
  {\bibinfo  {journal} {Phys. Rev. Lett.}\ }\textbf {\bibinfo {volume} {116}},\
  \bibinfo {pages} {063005} (\bibinfo {year} {2016})}\BibitemShut {NoStop}%
\bibitem [{\citenamefont {Kozyryev}\ \emph {et~al.}(2017)\citenamefont
  {Kozyryev}, \citenamefont {Baum}, \citenamefont {Matsuda}, \citenamefont
  {Augenbraun}, \citenamefont {Anderegg}, \citenamefont {Sedlack},\ and\
  \citenamefont {Doyle}}]{Kozyryev:17}%
  \BibitemOpen
  \bibfield  {author} {\bibinfo {author} {\bibfnamefont {I.}~\bibnamefont
  {Kozyryev}}, \bibinfo {author} {\bibfnamefont {L.}~\bibnamefont {Baum}},
  \bibinfo {author} {\bibfnamefont {K.}~\bibnamefont {Matsuda}}, \bibinfo
  {author} {\bibfnamefont {B.~L.}\ \bibnamefont {Augenbraun}}, \bibinfo
  {author} {\bibfnamefont {L.}~\bibnamefont {Anderegg}}, \bibinfo {author}
  {\bibfnamefont {A.~P.}\ \bibnamefont {Sedlack}},\ and\ \bibinfo {author}
  {\bibfnamefont {J.~M.}\ \bibnamefont {Doyle}},\ }\bibfield  {title} {\bibinfo
  {title} {Sisyphus laser cooling of a polyatomic molecule},\ }\href
  {https://doi.org/10.1103/PhysRevLett.118.173201} {\bibfield  {journal}
  {\bibinfo  {journal} {Phys. Rev. Lett.}\ }\textbf {\bibinfo {volume} {118}},\
  \bibinfo {pages} {173201} (\bibinfo {year} {2017})}\BibitemShut {NoStop}%
\bibitem [{\citenamefont {Mitra}\ \emph {et~al.}(2020)\citenamefont {Mitra},
  \citenamefont {Vilas}, \citenamefont {Hallas}, \citenamefont {Anderegg},
  \citenamefont {Augenbraun}, \citenamefont {Baum}, \citenamefont {Miller},
  \citenamefont {Raval},\ and\ \citenamefont {Doyle}}]{Mitra:20}%
  \BibitemOpen
  \bibfield  {author} {\bibinfo {author} {\bibfnamefont {D.}~\bibnamefont
  {Mitra}}, \bibinfo {author} {\bibfnamefont {N.~B.}\ \bibnamefont {Vilas}},
  \bibinfo {author} {\bibfnamefont {C.}~\bibnamefont {Hallas}}, \bibinfo
  {author} {\bibfnamefont {L.}~\bibnamefont {Anderegg}}, \bibinfo {author}
  {\bibfnamefont {B.~L.}\ \bibnamefont {Augenbraun}}, \bibinfo {author}
  {\bibfnamefont {L.}~\bibnamefont {Baum}}, \bibinfo {author} {\bibfnamefont
  {C.}~\bibnamefont {Miller}}, \bibinfo {author} {\bibfnamefont
  {S.}~\bibnamefont {Raval}},\ and\ \bibinfo {author} {\bibfnamefont {J.~M.}\
  \bibnamefont {Doyle}},\ }\bibfield  {title} {\bibinfo {title} {Direct laser
  cooling of a symmetric top molecule},\ }\href
  {https://doi.org/10.1126/science.abc5357} {\bibfield  {journal} {\bibinfo
  {journal} {Science}\ }\textbf {\bibinfo {volume} {369}},\ \bibinfo {pages}
  {1366} (\bibinfo {year} {2020})}\BibitemShut {NoStop}%
\bibitem [{\citenamefont {Augenbraun}\ \emph {et~al.}(2020)\citenamefont
  {Augenbraun}, \citenamefont {Doyle}, \citenamefont {Zelevinsky},\ and\
  \citenamefont {Kozyryev}}]{Augenbraun:20}%
  \BibitemOpen
  \bibfield  {author} {\bibinfo {author} {\bibfnamefont {B.~L.}\ \bibnamefont
  {Augenbraun}}, \bibinfo {author} {\bibfnamefont {J.~M.}\ \bibnamefont
  {Doyle}}, \bibinfo {author} {\bibfnamefont {T.}~\bibnamefont {Zelevinsky}},\
  and\ \bibinfo {author} {\bibfnamefont {I.}~\bibnamefont {Kozyryev}},\
  }\bibfield  {title} {\bibinfo {title} {Molecular asymmetry and optical
  cycling: Laser cooling asymmetric top molecules},\ }\href
  {https://doi.org/10.1103/PhysRevX.10.031022} {\bibfield  {journal} {\bibinfo
  {journal} {Phys. Rev. X}\ }\textbf {\bibinfo {volume} {10}},\ \bibinfo
  {pages} {031022} (\bibinfo {year} {2020})}\BibitemShut {NoStop}%
\bibitem [{\citenamefont {Vilas}\ \emph {et~al.}(2022)\citenamefont {Vilas},
  \citenamefont {Hallas}, \citenamefont {Anderegg}, \citenamefont {Robichaud},
  \citenamefont {Winnicki}, \citenamefont {Mitra},\ and\ \citenamefont
  {Doyle}}]{Vilas:22}%
  \BibitemOpen
  \bibfield  {author} {\bibinfo {author} {\bibfnamefont {N.~B.}\ \bibnamefont
  {Vilas}}, \bibinfo {author} {\bibfnamefont {C.}~\bibnamefont {Hallas}},
  \bibinfo {author} {\bibfnamefont {L.}~\bibnamefont {Anderegg}}, \bibinfo
  {author} {\bibfnamefont {P.}~\bibnamefont {Robichaud}}, \bibinfo {author}
  {\bibfnamefont {A.}~\bibnamefont {Winnicki}}, \bibinfo {author}
  {\bibfnamefont {D.}~\bibnamefont {Mitra}},\ and\ \bibinfo {author}
  {\bibfnamefont {J.~M.}\ \bibnamefont {Doyle}},\ }\bibfield  {title} {\bibinfo
  {title} {Magneto-optical trapping and sub-{Doppler} cooling of a polyatomic
  molecule},\ }\href {https://doi.org/10.1038/s41586-022-04620-5} {\bibfield
  {journal} {\bibinfo  {journal} {Nature}\ }\textbf {\bibinfo {volume} {606}},\
  \bibinfo {pages} {70} (\bibinfo {year} {2022})}\BibitemShut {NoStop}%
\bibitem [{\citenamefont {Zhang}\ \emph {et~al.}(2023)\citenamefont {Zhang},
  \citenamefont {Yu}, \citenamefont {Jadbabaie},\ and\ \citenamefont
  {Hutzler}}]{Zhang:23}%
  \BibitemOpen
  \bibfield  {author} {\bibinfo {author} {\bibfnamefont {C.}~\bibnamefont
  {Zhang}}, \bibinfo {author} {\bibfnamefont {P.}~\bibnamefont {Yu}}, \bibinfo
  {author} {\bibfnamefont {A.}~\bibnamefont {Jadbabaie}},\ and\ \bibinfo
  {author} {\bibfnamefont {N.~R.}\ \bibnamefont {Hutzler}},\ }\bibfield
  {title} {\bibinfo {title} {Quantum-enhanced metrology for molecular symmetry
  violation using decoherence-free subspaces},\ }\href
  {https://doi.org/10.1103/PhysRevLett.131.193602} {\bibfield  {journal}
  {\bibinfo  {journal} {Phys. Rev. Lett.}\ }\textbf {\bibinfo {volume} {131}},\
  \bibinfo {pages} {193602} (\bibinfo {year} {2023})}\BibitemShut {NoStop}%
\bibitem [{\citenamefont {Asnaashari}\ \emph {et~al.}(2023)\citenamefont
  {Asnaashari}, \citenamefont {Krems},\ and\ \citenamefont
  {Tscherbul}}]{Asnaashari:23}%
  \BibitemOpen
  \bibfield  {author} {\bibinfo {author} {\bibfnamefont {K.}~\bibnamefont
  {Asnaashari}}, \bibinfo {author} {\bibfnamefont {R.~V.}\ \bibnamefont
  {Krems}},\ and\ \bibinfo {author} {\bibfnamefont {T.~V.}\ \bibnamefont
  {Tscherbul}},\ }\bibfield  {title} {\bibinfo {title} {General classification
  of qubit encodings in ultracold diatomic molecules},\ }\href
  {https://doi.org/10.1021/acs.jpca.3c02835} {\bibfield  {journal} {\bibinfo
  {journal} {J. Phys. Chem. A}\ }\textbf {\bibinfo {volume} {127}},\ \bibinfo
  {pages} {6593} (\bibinfo {year} {2023})}\BibitemShut {NoStop}%
\bibitem [{\citenamefont {Zee}(2016)}]{ZeeBook}%
  \BibitemOpen
  \bibfield  {author} {\bibinfo {author} {\bibfnamefont {A.}~\bibnamefont
  {Zee}},\ }\href@noop {} {\emph {\bibinfo {title} {Group Theory in a Nutshell
  for Physicists}}},\ Vol.~\bibinfo {volume} {17}\ (\bibinfo  {publisher}
  {Princeton University Press},\ \bibinfo {year} {2016})\BibitemShut {NoStop}%
\bibitem [{\citenamefont {Runeson}\ and\ \citenamefont
  {Richardson}(2020)}]{Richarsdon:20}%
  \BibitemOpen
  \bibfield  {author} {\bibinfo {author} {\bibfnamefont {J.~E.}\ \bibnamefont
  {Runeson}}\ and\ \bibinfo {author} {\bibfnamefont {J.~O.}\ \bibnamefont
  {Richardson}},\ }\bibfield  {title} {\bibinfo {title} {Generalized spin
  mapping for quantum-classical dynamics},\ }\href@noop {} {\bibfield
  {journal} {\bibinfo  {journal} {J. Chem. Phys.}\ }\textbf {\bibinfo {volume}
  {152}},\ \bibinfo {pages} {084110} (\bibinfo {year} {2020})}\BibitemShut
  {NoStop}%
\bibitem [{\citenamefont {Hamley}\ \emph
  {et~al.}(2012{\natexlab{b}})\citenamefont {Hamley}, \citenamefont {Gerving},
  \citenamefont {Hoang}, \citenamefont {Bookjans},\ and\ \citenamefont
  {Chapman}}]{Chapman:12}%
  \BibitemOpen
  \bibfield  {author} {\bibinfo {author} {\bibfnamefont {C.~D.}\ \bibnamefont
  {Hamley}}, \bibinfo {author} {\bibfnamefont {C.}~\bibnamefont {Gerving}},
  \bibinfo {author} {\bibfnamefont {T.}~\bibnamefont {Hoang}}, \bibinfo
  {author} {\bibfnamefont {E.}~\bibnamefont {Bookjans}},\ and\ \bibinfo
  {author} {\bibfnamefont {M.~S.}\ \bibnamefont {Chapman}},\ }\bibfield
  {title} {\bibinfo {title} {Spin-nematic squeezed vacuum in a quantum gas},\
  }\href@noop {} {\bibfield  {journal} {\bibinfo  {journal} {Nat. Phys.}\
  }\textbf {\bibinfo {volume} {8}},\ \bibinfo {pages} {305} (\bibinfo {year}
  {2012}{\natexlab{b}})}\BibitemShut {NoStop}%
\bibitem [{\citenamefont {Pethick}\ and\ \citenamefont
  {Smith}(2008)}]{PethickBook}%
  \BibitemOpen
  \bibfield  {author} {\bibinfo {author} {\bibfnamefont {C.~J.}\ \bibnamefont
  {Pethick}}\ and\ \bibinfo {author} {\bibfnamefont {H.}~\bibnamefont
  {Smith}},\ }\href@noop {} {\emph {\bibinfo {title} {Bose--Einstein
  condensation in dilute gases}}}\ (\bibinfo  {publisher} {Cambridge university
  press},\ \bibinfo {year} {2008})\BibitemShut {NoStop}%
\end{thebibliography}%

\end{document}